\documentclass[conference]{IEEEtran}
\usepackage[boxruled]{algorithm2e}
\usepackage{cite}
\usepackage{graphicx}
\usepackage{psfrag}
\usepackage{subfigure}
\usepackage{url}
\usepackage{amsmath}
\usepackage{array}
\usepackage{amssymb}
\usepackage{amsfonts}
\usepackage{graphicx}
\usepackage{epstopdf}

\title{On Constellations for Physical Layer Network Coded Two-Way Relaying}
\begin{document}

\author{
Kiran Venugopal$^\ddagger$, Vishnu Namboodiri$^\dagger$ and B. Sundar Rajan$^\ddagger$\\
$^\dagger$Qualcomm India Private Limited, Hyderabad 500081, India. $^\ddagger$Dept. of ECE, IISc Bangalore 560012, India\\
Email: namboodiri.vishnu@gmail.com, \{kiran.v, bsrajan\}@ece.iisc.ernet.in\\
%\authorblockN{Kiran Venugopal}\\
%\authorblockA{Dept. of ECE, Indian Institute of Science \\
%Bangalore 560012, India\\
%Email: \\
%}
%\and
%\authorblockN{B. Sundar Rajan}
%\authorblockA{Dept. of ECE, Indian Institute of Science, \\Bangalore 560012, India\\
%Email: 
%}
}

\maketitle
\thispagestyle{empty}	
%\let\thefootnote\relax\footnote{Part of the content of this paper appears as an arXiv submission: 1301.4646v1 [cs.IT], 20 Jan 2013
%}
%\vspace{-0.15 in} 
%%%%%%%%
%%%%%%%%%%%%%%%%%%%%%%%%%%%%%%%%%%%%%%%%%%%%%%%%%%%%%%%%%%%%%%%%%%%%%%%%%%%%%%%%%%%%%
\begin{abstract}
Modulation schemes for two-way bidirectional relay network employing two phases: Multiple access (MA) phase and Broadcast (BC) phase and  using physical layer network coding are currently studied intensively. Recently, adaptive modulation schemes using Latin Squares to obtain network coding maps with the denoise and forward protocol have been reported with good end-to-end performance. These schemes work based on avoiding the detrimental effects of distance shortening in the effective receive constellation at the end of the MA phase at the relay. The channel fade states that create such distance shortening called  \textit{singular fade states}, are effectively removed using appropriate Latin squares. This scheme as well as all other known schemes studied so far use conventional regular PSK or QAM signal sets for the end users which lead to the relay using different sized constellations for the BC phase depending upon the fade state. In this work, we propose a 4-point signal set that would always require a 4-ary constellation for the BC phase for all the channel fade conditions. We also propose an 8-point constellation that gives better SER performance (gain of 1 dB) than 8-PSK while still using 8-ary constellation for BC phase like the case with 8-PSK. This is in spite of the fact that the proposed 8-point signal set has more number of singular fade states than for 8-PSK.
\end{abstract}
%\vspace{.1 in}
%%%%%%%%%%%%%%%%%%%%%%%%%%%%%%%%%%
\section{Background and Preliminaries}
\label{sec1}
\begin{figure}[htbp]
\centering
\subfigure[MA Phase]{
\includegraphics[totalheight=.25in,width=1.6in]{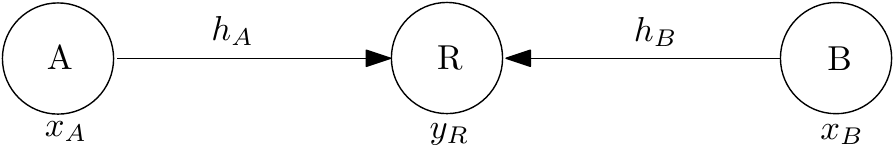}
\label{fig:phase1}
}
%\vspace{0.5 cm}
\subfigure[BC Phase]{
\includegraphics[totalheight=.25in,width=1.6in]{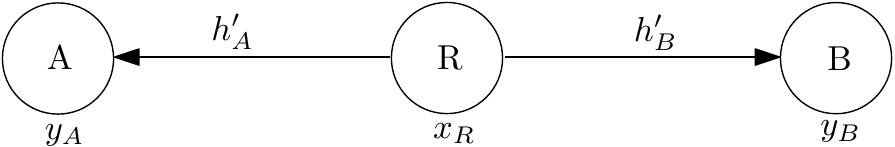}
\label{fig:phase2}
}
\caption{The Two Way Relay Channel}
\label{fig:relay_channel}
\end{figure}

%\vspace{-.1 in}
Physical layer network coding for wireless networks is a promising new area of research and a lot of work has been done in this area in recent times.  Wireless two-way relay network shown in Fig $\ref{fig:relay_channel}$, in which bidirectional data transfer takes place between the end nodes A and B with the help of the relay R is considered in this work. All the wireless links are assumed to be half-duplex. The De-noise and Forward (DNF) protocol for this case which consists of two phases: the \textit{multiple access} (MA) phase, during which both A and B simultaneously transmit to R and the \textit{broadcast} (BC) phase during which R transmits the network coded information to A and B is considered. Network coding map, also called the de-noising map, is used at R during the BC phase to broadcast a symbol to A and B in such a way that A (B) can decode the message of B (A), given that A (B) knows its own message. This scheme, however, results in Multiple Access Interference (MAI) at the relay node and strategies to mitigate MAI for two way relay channel were studied in \cite{PoY}. Design of coding schemes and network coding maps for uncoded transmission was studied in \cite{APT1}, and that for coded transmission was reported in \cite{APT2} and \cite{HeN}.

%\vspace{-0.4 in}
%%%%%%%%%%%%%%%%%%%%%%%%%%%%%%%%
%%%%%%%%%%%%%%%%%%%%%%%%%%%%%% 
It was observed in \cite{APT1} that for uncoded transmission, the network coding map used at the relay needs to be changed adaptively according to the channel fade coefficients, in order to optimize performance. Such Adaptive Network Coding (ANC) maps can be either obtained through computer search, as done in \cite{APT1} (called CNC algorithm), or analytically using  Latin Squares, as studied in \cite{NVR}, \cite{VNR} and \cite{NMRarX}. The Latin Square approach focuses on the removal of the harmful effects of a finite number of deep channel fade conditions, termed singular fade states, that can occur during the MA phase. It was also shown in \cite{NVR} that the Latin Square scheme gives better end-to-end performance than the CNC approach in \cite{APT1}.

Performance analysis of ANC schemes in \cite{APT1} and \cite{NVR} was done in \cite{VR}, where the connection between the removal of the harmful effect of singular states and the average end-to-end Symbol Error Rate (SER) performance was explained. Upper bounds were also given in \cite{VR} for SER in terms of the average error probability of point-to-point (SISO) fading channel and a term to account for the choice of network coding maps. From this, it is clear that by just using signal sets that give good performance in a SISO fading channel scenario, good end-to-end SER performance for wireless two-way relay network cannot be guaranteed. Also, it was observed in \cite{NMRarX} that using $M$-QAM instead of $M$-PSK at the end nodes gives a considerable improvement in performance. This suggests that the study of signal sets (possibly certain unconventional ones) for the MA phase assumes significance to ensure better end-to-end SER performance. However, to the best of our knowledge, there has been no prior work in this direction. %The only 4-point constellation considered in \cite{APT1}, \cite{NVR} and \cite{NMRarX} is the 4-PSK (same as 4-QAM with an average energy constraint) signal set.

\subsection{Signal Model}
Throughout, a quasi-static block fading scenario is assumed with perfect channel state information available at the receivers. All the fade coefficients are assumed to be Rician distributed with a Rician factor $K$. The fade coefficients associated with the A-R and B-R links are denoted  by $h_A$ and $h_B$ respectively and $h'_{A}$ and $h'_{B}$ denote the fading coefficients associated with the R-A and R-B links respectively. The ratio $h_B/h_A$ denoted as $z=\gamma e^{j \theta}$, where $\gamma \in \mathbb{R}^+$ and $-\pi \leq \theta < \pi$, is referred as the {\it fade state} and for simplicity, also denoted by $(\gamma, \theta)$.  All the additive noises are assumed to be $\mathcal{CN}(0,\sigma^2)$, which denotes the circularly symmetric complex Gaussian random variable with mean zero and variance $\sigma ^2$. 

Let $\mathcal{S}$ denote a signal set of size $M$ used at A and B. Assume that A (B) wants to transmit $x_A ~(x_B) \in {\mathcal{S}}$ to B (A). With $z_R$ denoting the additive noise at R, the received signal at $R$ during the MA phase is given by $y_R=h_{A} x_A + h_{B} x_B +z_R$. At the end of the MA phase, R evaluates $(\hat{x}_A^R,\hat{x}_B^R) \in \mathcal{S}^2$, which denotes the Maximum Likelihood (ML) estimate of $({x}_A,{x}_B)$ based on the received complex number $y_{R}$, i.e.,
\begin{align}
(\hat{x}_A^R,\hat{x}_B^R)=\arg\hspace{-0.5 cm}\min_{({x}'_A,{x}'_B) \in \mathcal{S}^2} \vert y_R-h_{A}{x}'_A-h_{B}{x}'_B\vert.
\end{align}

Depending on the value of $\gamma e^{j \theta}$, R chooses a map $\mathcal{M}^{\gamma,\theta}:\mathcal{S}^2 \rightarrow {\mathcal{S}'}(\gamma, \theta)$, where ${\mathcal{S}'}(\gamma, \theta)$ is the signal set (of size between $M$ and $M^2$) used by R during $BC$ phase and broadcasts $x_R=\mathcal{M}^{\gamma,\theta}(\hat{x}_A^R,\hat{x}_B^R)$. The received signals at A and B during the BC phase are respectively given by, $
y_A=h'_{A} x_R + z_A$ and $y_B=h'_{B} x_R + z_B$, where $z_A~(z_B)$ is the additive noise at A (B).

In order to ensure that A (B) is able to decode B's (A's) message, the map $\mathcal{M}^{\gamma,\theta}$ should satisfy the exclusive law \cite{APT1}, i.e.,
\begin{align*}
\mathcal{M}^{\gamma,\theta}(x_A,x_B) \neq \mathcal{M}^{\gamma,\theta}(x'_A,x_B), \; \mathrm{for} \;x_A \neq x'_A \; \mathrm{,} \; \forall x_B \in  \mathcal{S},\\
\mathcal{M}^{\gamma,\theta}(x_A,x_B) \neq \mathcal{M}^{\gamma,\theta}(x_A,x'_B), \; \mathrm{for} \;x_B \neq x'_B \; \mathrm{,} \;\forall x_A \in \mathcal{S}.
\end{align*}

Node A (B) can decode the message from B (A) by observing $y_A~(y_B)$ through ML decoding, since A (B) knows $x_A~(x_B)$ and ${\mathcal{M}}^{\gamma,\theta}$ satisfies the exclusive law.

\subsection{Removal of Singular Fade States by ANC}
\label{ANC_sfs}
Let $\mathcal{S}_{R}(\gamma,\theta)$ denote the effective constellation (normalized by $h_A$) seen at R, i.e., $
 \mathcal{S}_{R}(\gamma,\theta)=\left\lbrace x_i+\gamma e^{j \theta} x_j \vert x_i,x_j \in \mathcal{S}\right \rbrace$. The minimum distance between the points in $\mathcal{S}_{R}(\gamma,\theta)$ is given by \vspace{-0.2 in}

{\footnotesize
\begin{align}
\label{eqn_dmin} 
d_{min}(\gamma e^{j\theta})=\hspace{-0.5 cm}\min_{\substack {{(x_A,x_B),(x'_A,x'_B)}{{ \in \mathcal{S}^2 }} \\ {(x_A,x_B) \neq (x'_A,x'_B)}}}\hspace{-0.5 cm}\vert \left(x_A-x'_A\right)+\gamma e^{j \theta} \left(x_B-x'_B\right)\vert.
\end{align}} \vspace{-0.1 in}

Let $\mathcal{H}=\lbrace \gamma e^{j\theta} \in \mathbb{C} \vert d_{min}(\gamma e^{j \theta})=0 \rbrace$. The elements of $\mathcal{H}$ are called {\it the singular fade states} \cite{NVR}. The set $\mathcal{H}$ depends on the signal set used. The singular fade states $h \in {\mathcal{H}}$ are of the form
 \begin{equation}
\label{sing_expression}
h=\dfrac{x_A-x_A^{\prime}}{x_B^{\prime}-x_B} 
\end{equation}
 and is obtained by equating \eqref{eqn_dmin} to zero. With the difference constellation of $\mathcal{S}$ defined as $ \Delta\mathcal{S}= \left\lbrace x-x' : x,x' \in {\mathcal{S}} \right\rbrace $, the singular fade states can be written as $h=-d_k/d_l$, for some $d_k, d_l \in \Delta\mathcal{S}$.
 
The elements in $\mathcal{S}^2 $ which are mapped on to the same complex number in ${\mathcal{S}'}(\gamma, \theta)$ by the map $\mathcal{M}^{\gamma,\theta}$ are said to form a cluster. The set of all such clusters is denoted by $\mathcal{C}^{\gamma,\theta}$, to indicate that it is a function of $\gamma e^{j \theta}$. Since the map $\mathcal{M}^{\gamma,\theta}$ clusters the points in $\mathcal{S}^2 $, we interchangeably use the terms clustering and network coding map to denote $\mathcal{M}^{\gamma,\theta}$. The number of clusters for a given channel fade state $\gamma e^{j \theta}$, is the size of the set $\mathcal{C}^{\gamma,\theta}$ which is the same as $\vert {\mathcal{S}'} (\gamma, \theta) \vert $. The \textit{minimum cluster distance} of the clustering $\mathcal{C}$ is defined in \cite{NVR} as
{%\footnotesize
\begin{align*}
d_{min}^{\mathcal{C}}(\gamma e^{j \theta})=\hspace{-0.9 cm}\min_{\substack {{(x_A,x_B),(x'_A,x'_B)}{ \in \mathcal{S}^2,} \\ {\mathcal{M}^{\gamma,\theta}(x_A,x_B) \neq \mathcal{M}^{\gamma,\theta}(x'_A,x'_B)}}}\hspace{-0.8 cm}\vert \left( x_A-x'_A\right)+\gamma e^{j \theta} \left(x_B-x'_B\right)\vert.
\end{align*}}

A clustering $\mathcal{C}$ is said to remove a singular fade state $ h \in \mathcal{H}$, if $d_{min}^{\mathcal{C}}(h)>0$.  For a singular fade state $h \in \mathcal{H}$, let $\mathcal{C}^{\lbrace h\rbrace}$ denote a clustering which removes the singular fade state $h$, and ${\mathcal{S}'} (h)$ denote the corresponding signal set used.

The relay clusterings satisfying the exclusive law form Latin Squares \cite{NVR}. The $(i,j)^{th}$ entry of the Latin square (corresponding to the singular fade state $h$) is the index of the signal point from ${\mathcal{S}'} (h)$ that is transmitted by R during the BC phase where the indices of $\hat{x}_A^R$ and $\hat{x}_B^R$ $\in {\mathcal{S}}$ are $i$ and $j$ respectively. Let $t(h)$ denote the value of the largest entry in the Latin square of $h$. 

For channel fade states other than the singular fade states, among all the network maps which remove the singular fade states, the one which optimizes the performance is chosen. Since the network maps which remove the singular fade states are known to all the three nodes and are finite in number, the clustering used for a particular realization of the fade state can be indicated by R to A and B using overhead bits.

%%%%%%%%%%%%%%%%%%%%%%%%%%%%%%%%%%%%%%%%%%%%%%%%
The main contributions of this paper are as follows:
\begin{itemize}
\item The desirable features of the $M$-ary signal set $\mathcal{S}$ (for a general $M$) used in the MA phase for wireless bi-directional two-way relay network to ensure good performance are identified to be the following
\begin{itemize}
\item large value for the minimum Euclidean distance of $\mathcal{S}$.
\item minimum number of clusterings for each singular fade state $h$.
\end{itemize} 
\item It is shown that signal sets having fewer number of singular fade states do not invariably give better end-to-end SER performance; examples of signal sets having larger number of singular fade states and yet giving better end-to-end SER performance are given. 
\item In \cite{APT1} and \cite{NVR} it was noted that when QPSK is used for MA phase, the BC phase required a 5-ary constellation for certain channel conditions. We identify an alternate 4-ary constellation for the MA phase which requires only 4-ary signal set for all channel conditions. This constellation is also shown to give better end-to-end SER performance than 4-PSK with the requirement that only 4-ary constellation is used by the relay.
\item An 8-ary signal set is proposed as an alternative for the existing 8-PSK signal set, which results in better end-to-end SER performance.
\end{itemize}

The remaining content is organized as follows: In Section \ref{sec2} we briefly summarize the factors determining the performance of physical layer network coding schemes for two-way relay network and list out the desirable features for the choice of the constellation used for MA phase. In Section \ref{sec3}, an alternate 4-point signal set which gives clear advantages over 4-PSK is presented. An alternate 8-point signal set that gives better SER performance than 8-PSK is given in Section \ref{sec4}. Simulation results are given in Section \ref{sec5}.
%%%%%%%%%%%%%%%%%%%%%%%%%%%%%%%%%%%%%%%%%%%%%%%%%%%%%%%%%%%%%%%%%%%%%%%%%%%%%%%%%%%%%
%\vspace{.1 in}
%%%%%%%%%%%%%%%%%%%%%%%%%%%%%%%%%%
\section{Performance determining factors}
\label{sec2}
In this section, we summarize the error analysis results for wireless two-way relaying scenario described in \cite{VR}. Following this, the desirable features in the signal sets used for MA phase are described.

\subsection{Average end-to-end SER}
Let ${\hat{x}}_B^A({\hat{x}}_A^B)$ denote the ML estimate of $x_B(x_A)$ at node A(B) at the end of the BC phase. For a given channel realisation \mbox{$H=(h_A, h_B, h'_A, h'_B)$}, the error event \mbox{${\hat{x}}_B^A \neq x_B$} can be upper bounded by the sum of two conditional error events viz., one occurring due to the relay R mapping $({\hat{x}_A},{\hat{x}_B})$ to a wrong cluster (termed cluster error) and hence transmitting a wrong symbol during BC phase and the second occurring in spite of the fact that R sent the correct symbol. The first term is denoted by $P_H^{CE}$ and the second by $P^{A,BC}_H$. Taking expectation over $H$, the end-to-end average SER is upper bounded by the sum of the average Cluster Error Probability (CEP), $P^{CE}$ and the average probability that an error occurs during BC phase given that no cluster error occurred at R, denoted as $P^{A,BC}$. 

With $P^{pp}(\mathcal{S})$ denoting the average SER of the point-to-point fading channel using the signal set $\mathcal{S}$, it was shown in \cite{VR} that a tight upper bound for $P^{CE}$ at high SNR values is equal to $2P^{pp}(\mathcal{S})$. The signal sets used for BC phase determine $P^{A,BC}$. Let $\lambda_h$ denote the fraction of time the signal set $\mathcal{S'}(h)$ is used in the BC phase. Then $P^{A,BC}=\Sigma_{h \in {\mathcal{H}}} \lambda_h P^{pp}(\mathcal{S'}(h))$. 

\subsection{Desirable features for selecting the signal set $\mathcal{S}$}

With an average energy constraint $E$, the factor that determines $P^{pp}(\mathcal{S})$ at high SNR values for a signal set $\mathcal{S}$ is the minimum Euclidean distance $d_{min}({\mathcal{S}})$ of the constellation. However, choosing signal set for MA phase that gives smaller values for $P^{pp}(\mathcal{S})$ need not necessarily ensure better end-to-end SER performance for the two-way relay network. This is because, such a signal set may require more number of clusterings (hence requiring larger sized ${\mathcal{S'}}(h)$) for some singular fade state $h$. This in turn increases the cluster error probability in addition to degrading the performance in the BC phase through the factor $P^{pp}(\mathcal{S'}(h))$. Note that the minimum value for $|\mathcal{S'}(h)|$ is $M$. The number of singular fade states, given by $\vert {\mathcal{H}} \vert$, as such does not play any role in determining the end-to-end SER performance as long as the clusterings remove all the singular fade states. However, a large value for $\vert {\mathcal{H}} \vert$ would require the use of more number of overhead bits to indicate the choice of the network coding map to the end nodes A and B. We summarize the desirable features for the choice of the signal set $\mathcal{S}$ used by the end nodes during the MA phase as follows
\begin{itemize}
\item Larger value for $d_{min}({\mathcal{S}})$ is preferred so as to get improved performance in MA phase. This ensures that the average Cluster Error Probability is reduced at the relay.
\item The number of clusterings for each $h \in {\mathcal{H}}$ must be minimal i.e., $M$ itself. With this, the average probability that an error occurs during the BC phase, assuming the relay transmitted the correct network coded information, is minimized.
\end{itemize}
With these features satisfied, we come up with two constellations, one for \mbox{$M=4$} and another for \mbox{$M=8$} in the next two sections.

\section{An Alternate 4-point Constellation}
\label{sec3}
The only 4-ary constellation that has been considered in the context of physical layer network coded two-way relaying in \cite{APT1}, \cite{NVR} and \cite{NMRarX} is the conventional 4-PSK (same as 4-QAM with an energy constraint). In this section we present an alternate constellation with 4 points that gives advantages over 4-PSK as described below.

\subsection{Issues with 4-PSK signal set}

It was noted in \cite{APT1} and \cite{NVR} that when 4-PSK is used for MA phase, in order to overcome MAI at relay R, for several clusterings (8 out of the total 12 clusterings), an unconventional 5-point signal set has to be used for the BC phase. After channel quantization, both \cite{APT1} and \cite{NVR} identify regions in the complex plane corresponding to values of $\gamma e^{j\theta}$ where the BC phase requires a 5-point constellation. In particular, this scenario arose in \cite{NVR} because certain singular fade states $h$ for 4-PSK needed Latin Squares with \mbox{$t(h)=5$}. For the proposed 4-point constellation, the Latin Squares always have \mbox{$t=4$} and hence the same 4-point constellation can be used at the relay. In particular, the use of the unconventional 5-point signal set can be totally avoided.

\begin{figure}[htbp]
\centering
\includegraphics[totalheight=2in,width=2in]{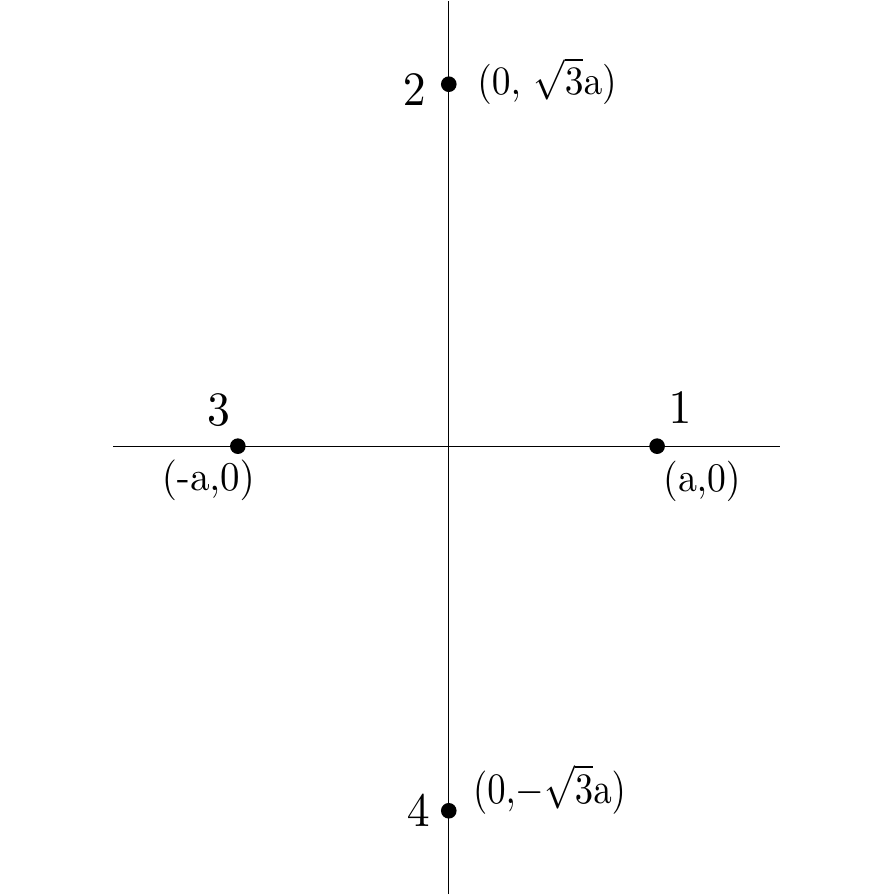} 
\caption{The proposed 4-point constellation ${\mathcal{S}}_{4}$}
\label{fig:proposed4}
\end{figure}

The alternate 4-point constellation, denoted by $\mathcal{S}_4$ is shown in Fig. \ref{fig:proposed4}. This constellation appears in \cite{FGLLQ} in the context of band-limited point-to-point AWGN channel. In Fig. \ref{fig:proposed4}, the quantity $a$ is chosen so as to meet an average energy constraint $E$. 

\begin{figure}[htbp]
\centering
\includegraphics[totalheight=2in,width=2in]{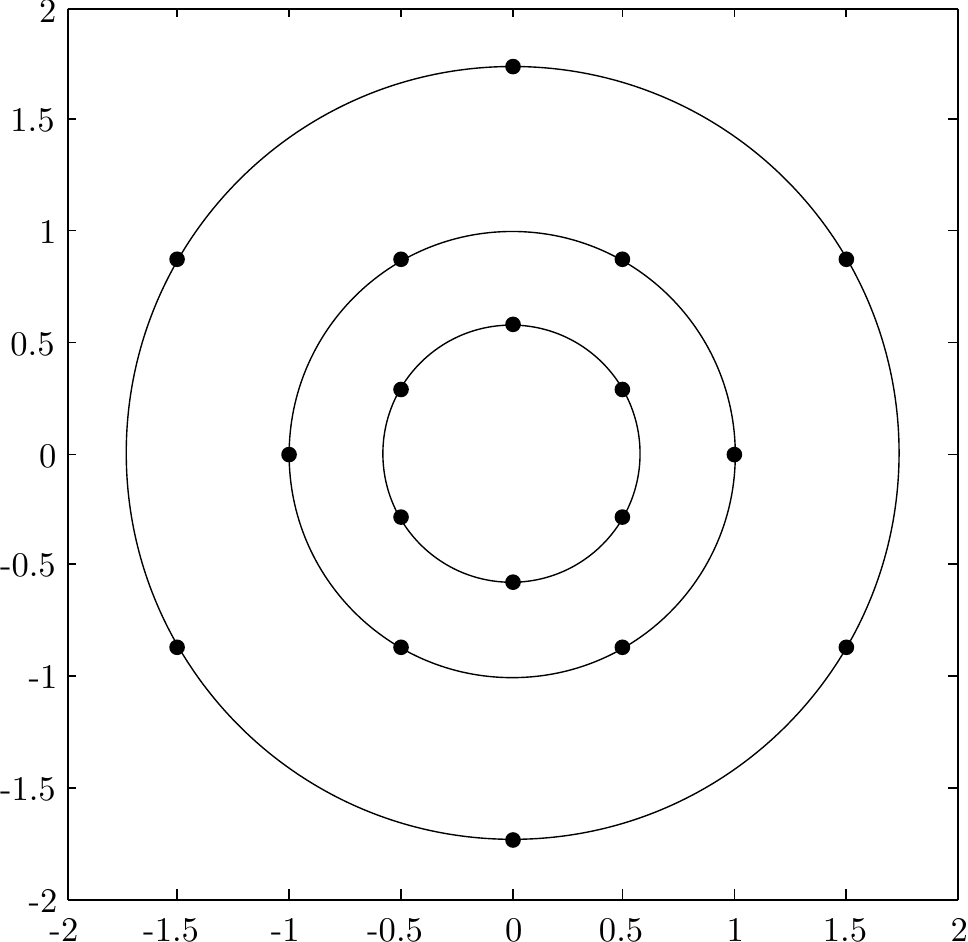} 
\caption{Singular fade states of ${\mathcal{S}}_{4}$}
\label{fig:sing_proposed4}
\end{figure}

The geometry of the signal set is such that the sets of points $\{1, 2, 3 \}$ and $\{1, 3, 4 \}$ (in Fig. \ref{fig:proposed4}) form vertices of an equilateral triangle. The constellation has 18 singular fade states viz., $\left \{(\pm 0.5 \pm j\frac{\sqrt{3}}{2}),\hspace{0.1 cm} \pm 1,\hspace{0.08 cm} \pm j \sqrt{3},\hspace{0.08 cm} (\pm 1.5 \pm j\frac{\sqrt{3}}{2}),\hspace{0.08 cm} \frac{2}{\pm 3 \pm j\sqrt{3}},\hspace{0.08 cm} \frac{\pm 1}{ j\sqrt{3}} \right \}$ which is more than that for 4-PSK (12 singular fade states). The singular fade states of ${\mathcal{S}}_{4}$ lie on three concentric circles centered at the origin and of radii $\{1, \sqrt{3}, \frac{1}{\sqrt{3}} \}$ as shown in Fig. \ref{fig:sing_proposed4}. Table \ref{4_point_ compare} summarizes the key aspects of the 4-PSK and ${\mathcal{S}}_{4}$ which are relevant in our context.

\begin{table}[htbp]
\centering
\caption{Comparison between 4-PSK and ${\mathcal{S}}_{4}$}
\label{4_point_ compare}
\begin{tabular}{|c|c|c|}
\hline Feature & 4-PSK & ${\mathcal{S}}_{4}$ \\
\hline $d_{min}$ (for \mbox{$E=1$}) &  $\sqrt{2}$ & $\sqrt{2}$ \\
\hline Number of  &  & \\
 singular fade states & 12 & 18\\
\hline Max. size of constellation &  & \\
 required for BC phase & 5 &4 \\
\hline
\end{tabular}
\end{table}

Next we provide the Latin Squares for ${\mathcal{S}}_{4}$ in Fig. \ref{tab:LS_4} to show that no singular fade state $h$ of ${\mathcal{S}}_{4}$ has $t(h) > 4$. This ensures that the relay R can always use a 4-point signal set during the BC phase. The Latin squares of the singular fade states lying on the circle with radius $\frac{1}{\sqrt{3}}$ can be obtained by taking the transpose of the corresponding singular fade states (inverses) lying on the circle with radius $\sqrt{3}$,  and from the Latin Square corresponding to a singular fade state $h$, those corresponding to $-h$ and $h^*$ (conjugate of $h$) can be obtained using row/column permutations. This is because the signal set ${\mathcal{S}}_{4}$ is such that if $s \in {\mathcal{S}}_{4}$, then $-s, s^* \in {\mathcal{S}}_{4}$. Also the same Latin square (XOR network coding map) can remove the singular fade states \mbox{$h=1$} and \mbox{$h=j \sqrt{3}$}. So from the 3 Latin Squares shown in Fig. \ref{tab:LS_4}, the Latin Squares for all the remaining 12 singular fade states can be derived using row and/or column permutation(s).

For the purpose of comparison with 4-PSK, notice that for 4-PSK, the clusterings which require 4-ary signal set for BC phase can remove only the four singular fade states $\{ \pm 1, \pm j \}$. The remaining 8 singular fade states require the relay to invariably use 5-ary signal set for BC phase. Simulation results in Section \ref{sec5} show that when we impose an additional constraint that the relay can use only 4-ary constellation for the BC phase in the physical layer network coded two-way relay scenario (thus only removing 4 singular fade states for 4-PSK), the SER performance of ${\mathcal{S}}_{4}$ is better than that of 4-PSK. 

\begin{figure}[htbp]
\centering
%\subfigure[$h=-0.5-j\frac{\sqrt{3}}{2}$]{
%\begin{tabular}{|c|c|c|c|}
%\hline 1 & 4 & 2 & 3\\
%\hline 3 & 2 & 4 & 1\\
%\hline 4 & 1 & 3 & 2\\
%\hline 2 & 3 & 1 & 4\\
%\hline
%\end{tabular}
%}
%\subfigure[$h=0.5-j\frac{\sqrt{3}}{2}$]{
%\begin{tabular}{|c|c|c|c|}
%\hline 1 & 4 & 2 & 3\\
%\hline 2 & 3 & 1 & 4\\
%\hline 3 & 2 & 4 & 1\\
%\hline 4 & 1 & 3 & 2\\
%\hline
%\end{tabular}
%}
\subfigure[$h=0.5+j\frac{\sqrt{3}}{2}$]{
\begin{tabular}{|c|c|c|c|}
\hline 4 & 1 & 3 & 2\\
\hline 2 & 3 & 1 & 4\\
\hline 1 & 4 & 2 & 3\\
\hline 3 & 2 & 4 & 1\\
\hline
\end{tabular}
}
%\subfigure[$h=-0.5+j\frac{\sqrt{3}}{2}$]{
%\begin{tabular}{|c|c|c|c|}
%\hline 3 & 2 & 4 & 1\\
%\hline 4 & 1 & 3 & 2\\
%\hline 1 & 4 & 2 & 3\\
%\hline 2 & 3 & 1 & 4\\
%\hline
%\end{tabular}
%}
\subfigure[$h=1$, \mbox{$h=j \sqrt{3}$}]{
\begin{tabular}{|c|c|c|c|}
\hline 1 & 2 & 3 & 4\\
\hline 2 & 1 & 4 & 3\\
\hline 3 & 4 & 1 & 2\\
\hline 4 & 3 & 2 & 1\\
\hline
\end{tabular}
}
\subfigure[$h=1.5+j\frac{\sqrt{3}}{2}$]{
\begin{tabular}{|c|c|c|c|}
\hline 4 & 3 & 2 & 1\\
\hline 3 & 4 & 1 & 2\\
\hline 1 & 2 & 4 & 3\\
\hline 2 & 1 & 3 & 4\\
\hline
\end{tabular}
}

\caption{Latin squares for 3 singular states of ${\mathcal{S}}_{4}$ from which all the remaining Latin squares can be derived}
\label{tab:LS_4}
\end{figure}

%\begin{figure}[htbp]
%\centering
%\subfigure[$h=-1.5-j\frac{\sqrt{3}}{2}$]{
%\begin{tabular}{|c|c|c|c|}
%\hline 2 & 1 & 3 & 4\\
%\hline 1 & 2 & 4 & 3\\
%\hline 3 & 4 & 1 & 2\\
%\hline 4 & 3 & 2 & 1\\
%\hline
%\end{tabular}
%}
%\subfigure[$h=1.5-j\frac{\sqrt{3}}{2}$]{
%\begin{tabular}{|c|c|c|c|}
%\hline 4 & 2 & 1 & 3\\
%\hline 1 & 4 & 3 & 2\\
%\hline 2 & 3 & 4 & 1\\
%\hline 3 & 1 & 2 & 4\\
%\hline
%\end{tabular}
%}
%\subfigure[$h=1.5+j\frac{\sqrt{3}}{2}$]{
%\begin{tabular}{|c|c|c|c|}
%\hline 4 & 3 & 2 & 1\\
%\hline 3 & 4 & 1 & 2\\
%\hline 1 & 2 & 4 & 3\\
%\hline 2 & 1 & 3 & 4\\
%\hline
%\end{tabular}
%}
%\subfigure[$h=-1.5+j\frac{\sqrt{3}}{2}$]{
%\begin{tabular}{|c|c|c|c|}
%\hline 4 & 2 & 3 & 1\\
%\hline 3 & 1 & 2 & 4\\
%\hline 2 & 4 & 1 & 3\\
%\hline 1 & 3 & 4 & 2\\
%\hline
%\end{tabular}
%}
%\subfigure[$h=-j\sqrt{3}$]{
%\begin{tabular}{|c|c|c|c|}
%\hline 4 & 3 & 2 & 1\\
%\hline 1 & 2 & 3 & 4\\
%\hline 3 & 1 & 4 & 2\\
%\hline 2 & 4 & 1 & 3\\
%\hline
%\end{tabular}
%}
%\subfigure[$h=j\sqrt{3}$]{
%\begin{tabular}{|c|c|c|c|}
%\hline 4 & 3 & 2 & 1\\
%\hline 3 & 4 & 1 & 2\\
%\hline 2 & 1 & 4 & 3\\
%\hline 1 & 2 & 3 & 4\\
%\hline
%\end{tabular}
%}
%\caption{Latin squares for singular states with $|h|=\sqrt{3}$ for \mbox{${\mathcal{S}}_{4-proposed}$} }
%\label{tab:LS>1}
%\end{figure}

\section{The proposed 8-point signal set}
\label{sec4}
\begin{figure}[ht]
\centering
\includegraphics[totalheight=2in,width=2in]{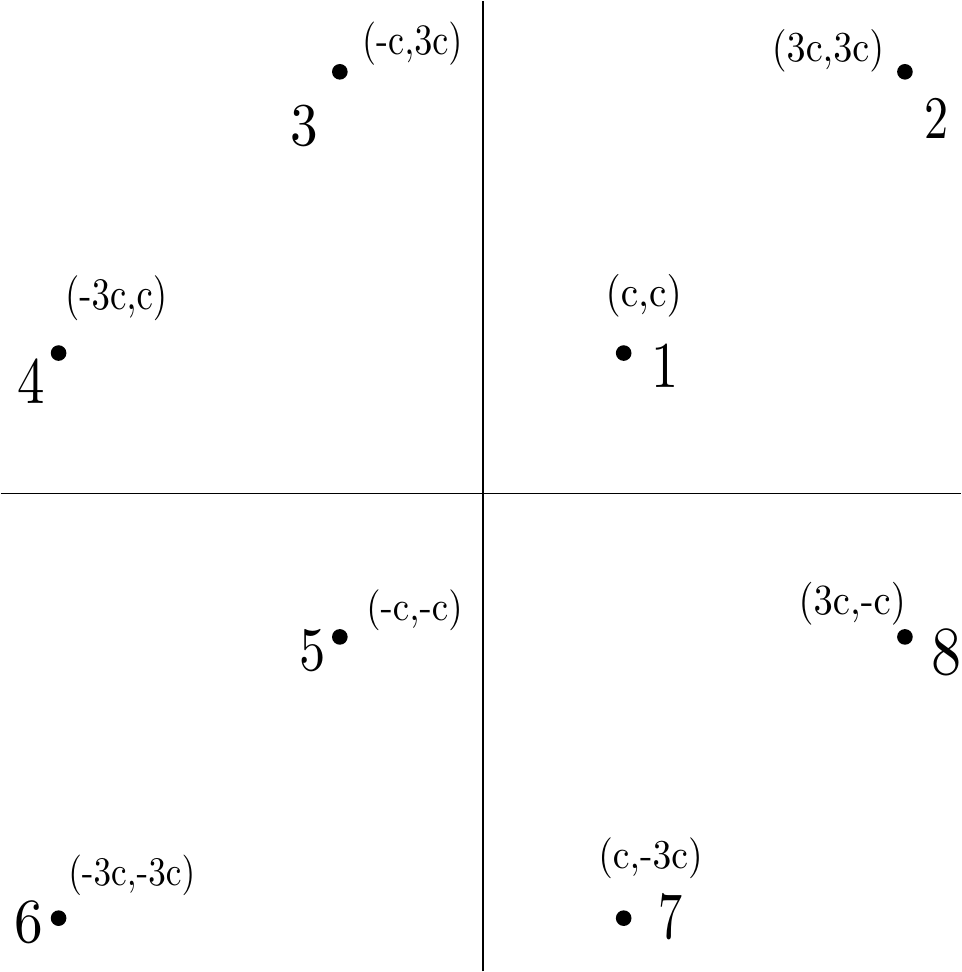} 
\caption{The 8-cross QAM constellation}
\label{fig:cross8}
\end{figure}

\begin{figure}[ht]
\centering
\includegraphics[totalheight=2in,width=2in]{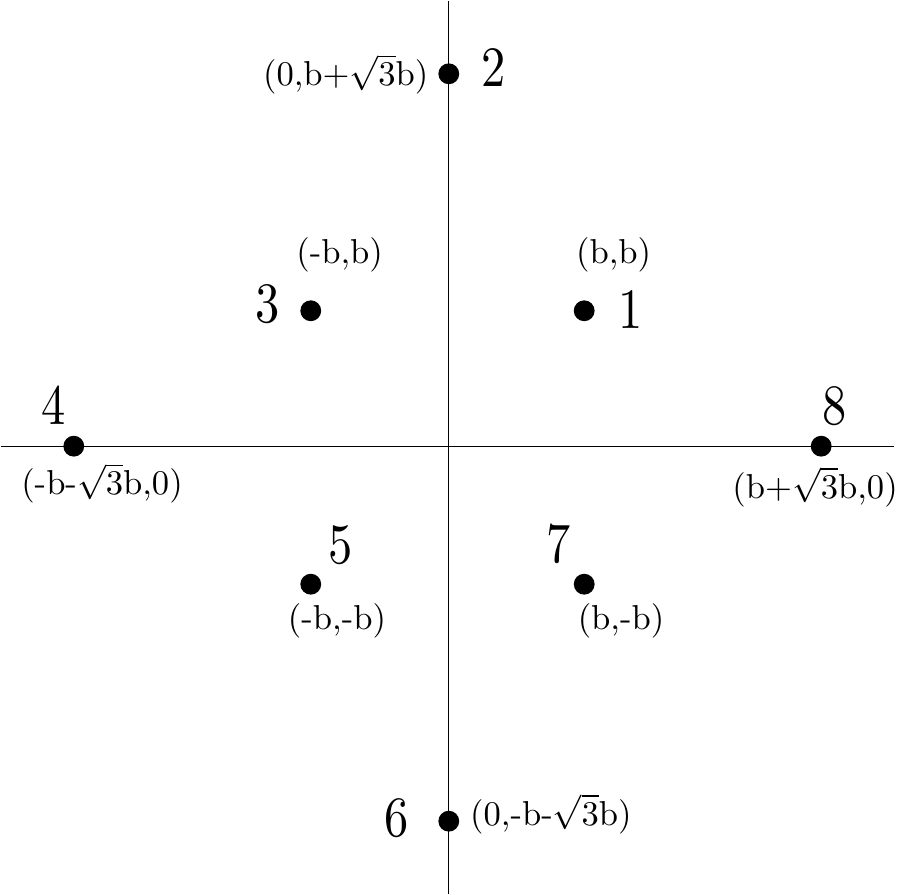} 
\caption{The proposed 8 point constellation ${\mathcal{S}}_{8}$}
\label{fig:proposed8}
\end{figure}

In this section we consider the case when $M$=8 point signal set is used by the end users. The conventional 8-PSK was studied in \cite{NVR} and was shown to require only 8 point constellation for BC phase for all channel realisations. We compare this with the 8-cross QAM constellation (shown in Fig. \ref{fig:cross8}) and propose an 8-point constellation which is shown to give the best performance in terms of the average end-to-end SER performance for $M$=8. The proposed 8-point constellation is shown in Fig. \ref{fig:proposed8} and is denoted by ${\mathcal{S}}_{8}$ . The advantage of 8-PSK that it always requires the relay R to use 8-point signal set for the BC phase carries forward to ${\mathcal{S}}_{8}$ also. While the number of singular fade states for 8-PSK is 104, that for ${\mathcal{S}}_{8}$ is 108. The key features of all the three above mentioned constellations are summarized in Table \ref{8_point_ compare}. From Table \ref{8_point_ compare} we can see that while all the three signal sets require only 8 point constellations during the BC phase, ${\mathcal{S}}_{8}$ has the largest value for $d_{min}$ and hence gives the best end-to-end SER performance. Simulation results in Section \ref{sec5} validate this.

\begin{table}[htbp]
\centering
\caption{Comparison between 8-PSK, 8-cross QAM and ${\mathcal{S}}_{8}$}
\label{8_point_ compare}
\begin{tabular}{|c|c|c|c|}
\hline Feature & 8-PSK & 8-cross QAM & ${\mathcal{S}}_{8}$ \\
\hline $d_{min}$ (for \mbox{$E=1$}) &  0.7653 & 0.8944 & 0.9194 \\
\hline Number of  &  &  &  \\
 singular fade states & 104 &  116 & 108 \\
\hline Max. size of constellation &  &  &  \\
 required for BC phase &8 & 8& 8\\
\hline
\end{tabular}
\end{table} 

\begin{figure}[ht]
\centering
\includegraphics[totalheight=2in,width=2in]{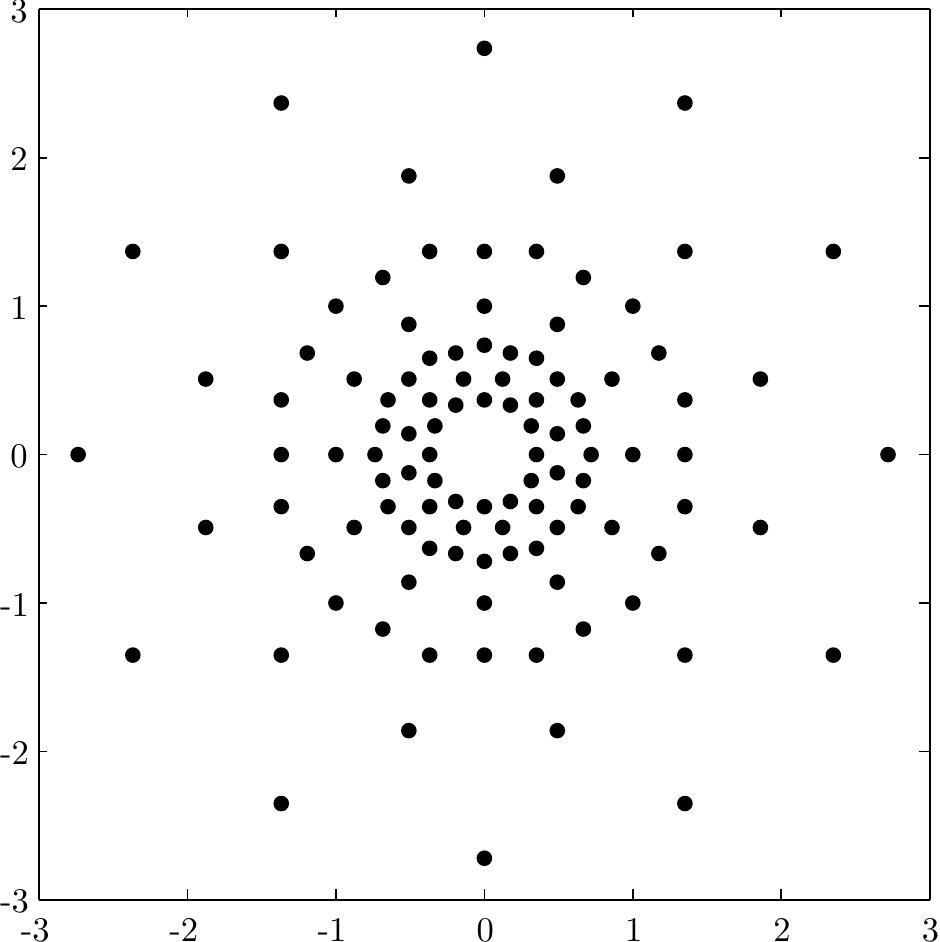} 
\caption{Singular fade states of ${\mathcal{S}}_{8}$}
\label{fig:sing_proposed8}
\end{figure}

From Fig. \ref{fig:proposed8}, points $\{1,3,5,7\}$ form QPSK signal points and the sets of points $\{1,2,3 \}$, $\{3,4,5 \}$, $\{5,6,7 \}$ and $\{7,8,1 \}$ form the vertices of an equilateral triangle. This constellation finds mention in \cite{Proak}. The 108 singular fade states for ${\mathcal{S}}_{8}$ are distributed in the complex fade state space as shown in Fig. \ref{fig:sing_proposed8}.

\begin{figure*}[htbp]
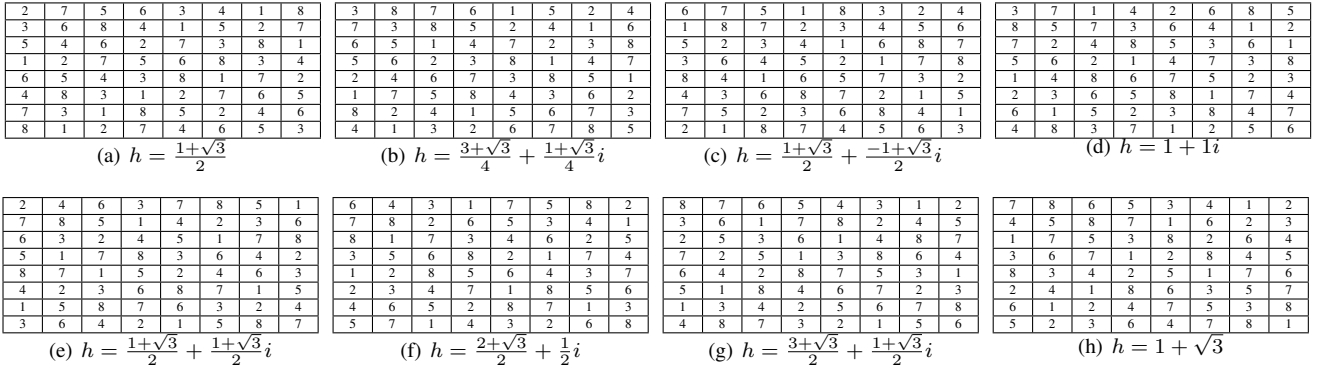
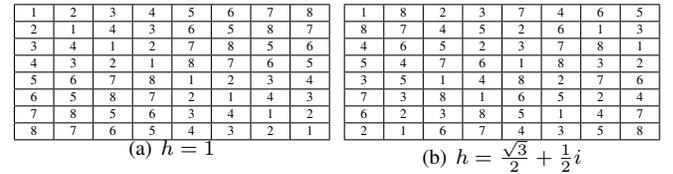

\centering
\tiny{
\subfigure[$h=\frac{1+\sqrt{3}}{2} $]{
\begin{tabular}{|c|c|c|c|c|c|c|c|}
\hline 2 & 7 & 5 & 6 & 3 & 4 & 1 & 8\\
\hline 3 & 6 & 8 & 4 & 1 & 5 & 2 & 7\\
\hline 5 & 4 & 6 & 2 & 7 & 3 & 8 & 1\\
\hline 1 & 2 & 7 & 5 & 6 & 8 & 3 & 4\\
\hline 6 & 5 & 4 & 3 & 8 & 1 & 7 & 2\\
\hline 4 & 8 & 3 & 1 & 2 & 7 & 6 & 5\\
\hline 7 & 3 & 1 & 8 & 5 & 2 & 4 & 6\\
\hline 8 & 1 & 2 & 7 & 4 & 6 & 5 & 3\\
\hline
\end{tabular}
}
\subfigure[$h=\frac{3+\sqrt{3}}{4} + \frac{1+\sqrt{3}}{4}i$]{
\begin{tabular}{|c|c|c|c|c|c|c|c|}
\hline 3 & 8 & 7 & 6 & 1 & 5 & 2 & 4\\
\hline 7 & 3 & 8 & 5 & 2 & 4 & 1 & 6\\
\hline 6 & 5 & 1 & 4 & 7 & 2 & 3 & 8\\
\hline 5 & 6 & 2 & 3 & 8 & 1 & 4 & 7\\
\hline 2 & 4 & 6 & 7 & 3 & 8 & 5 & 1\\
\hline 1 & 7 & 5 & 8 & 4 & 3 & 6 & 2\\
\hline 8 & 2 & 4 & 1 & 5 & 6 & 7 & 3\\
\hline 4 & 1 & 3 & 2 & 6 & 7 & 8 & 5\\
\hline
\end{tabular}
}
\subfigure[$h=\frac{1+\sqrt{3}}{2} + \frac{-1+\sqrt{3}}{2}i$]{
\begin{tabular}{|c|c|c|c|c|c|c|c|}
\hline 6 & 7 & 5 & 1 & 8 & 3 & 2 & 4\\
\hline 1 & 8 & 7 & 2 & 3 & 4 & 5 & 6\\
\hline 5 & 2 & 3 & 4 & 1 & 6 & 8 & 7\\
\hline 3 & 6 & 4 & 5 & 2 & 1 & 7 & 8\\
\hline 8 & 4 & 1 & 6 & 5 & 7 & 3 & 2\\
\hline 4 & 3 & 6 & 8 & 7 & 2 & 1 & 5\\
\hline 7 & 5 & 2 & 3 & 6 & 8 & 4 & 1\\
\hline 2 & 1 & 8 & 7 & 4 & 5 & 6 & 3\\
\hline
\end{tabular}
}
\subfigure[$h=1 + 1i$]{
\begin{tabular}{|c|c|c|c|c|c|c|c|}
\hline 3 & 7 & 1 & 4 & 2 & 6 & 8 & 5\\
\hline 8 & 5 & 7 & 3 & 6 & 4 & 1 & 2\\
\hline 7 & 2 & 4 & 8 & 5 & 3 & 6 & 1\\
\hline 5 & 6 & 2 & 1 & 4 & 7 & 3 & 8\\
\hline 1 & 4 & 8 & 6 & 7 & 5 & 2 & 3\\
\hline 2 & 3 & 6 & 5 & 8 & 1 & 7 & 4\\
\hline 6 & 1 & 5 & 2 & 3 & 8 & 4 & 7\\
\hline 4 & 8 & 3 & 7 & 1 & 2 & 5 & 6\\
\hline
\end{tabular}
}
\subfigure[$h=\frac{1+\sqrt{3}}{2} + \frac{1+\sqrt{3}}{2}i$]{
\begin{tabular}{|c|c|c|c|c|c|c|c|}
\hline 2 & 4 & 6 & 3 & 7 & 8 & 5 & 1\\
\hline 7 & 8 & 5 & 1 & 4 & 2 & 3 & 6\\
\hline 6 & 3 & 2 & 4 & 5 & 1 & 7 & 8\\
\hline 5 & 1 & 7 & 8 & 3 & 6 & 4 & 2\\
\hline 8 & 7 & 1 & 5 & 2 & 4 & 6 & 3\\
\hline 4 & 2 & 3 & 6 & 8 & 7 & 1 & 5\\
\hline 1 & 5 & 8 & 7 & 6 & 3 & 2 & 4\\
\hline 3 & 6 & 4 & 2 & 1 & 5 & 8 & 7\\
\hline
\end{tabular}
}
\subfigure[$h=\frac{2+\sqrt{3}}{2} + \frac{1}{2}i$]{
\begin{tabular}{|c|c|c|c|c|c|c|c|}
\hline 6 & 4 & 3 & 1 & 7 & 5 & 8 & 2\\
\hline 7 & 8 & 2 & 6 & 5 & 3 & 4 & 1\\
\hline 8 & 1 & 7 & 3 & 4 & 6 & 2 & 5\\
\hline 3 & 5 & 6 & 8 & 2 & 1 & 7 & 4\\
\hline 1 & 2 & 8 & 5 & 6 & 4 & 3 & 7\\
\hline 2 & 3 & 4 & 7 & 1 & 8 & 5 & 6\\
\hline 4 & 6 & 5 & 2 & 8 & 7 & 1 & 3\\
\hline 5 & 7 & 1 & 4 & 3 & 2 & 6 & 8\\
\hline
\end{tabular}
}
\subfigure[$h=\frac{3+\sqrt{3}}{2} + \frac{1+\sqrt{3}}{2}i$]{
\begin{tabular}{|c|c|c|c|c|c|c|c|}
\hline 8 & 7 & 6 & 5 & 4 & 3 & 1 & 2\\
\hline 3 & 6 & 1 & 7 & 8 & 2 & 4 & 5\\
\hline 2 & 5 & 3 & 6 & 1 & 4 & 8 & 7\\
\hline 7 & 2 & 5 & 1 & 3 & 8 & 6 & 4\\
\hline 6 & 4 & 2 & 8 & 7 & 5 & 3 & 1\\
\hline 5 & 1 & 8 & 4 & 6 & 7 & 2 & 3\\
\hline 1 & 3 & 4 & 2 & 5 & 6 & 7 & 8\\
\hline 4 & 8 & 7 & 3 & 2 & 1 & 5 & 6\\
\hline
\end{tabular}
}
\subfigure[$h=1+\sqrt{3} $]{
\begin{tabular}{|c|c|c|c|c|c|c|c|}
\hline 7 & 8 & 6 & 5 & 3 & 4 & 1 & 2\\
\hline 4 & 5 & 8 & 7 & 1 & 6 & 2 & 3\\
\hline 1 & 7 & 5 & 3 & 8 & 2 & 6 & 4\\
\hline 3 & 6 & 7 & 1 & 2 & 8 & 4 & 5\\
\hline 8 & 3 & 4 & 2 & 5 & 1 & 7 & 6\\
\hline 2 & 4 & 1 & 8 & 6 & 3 & 5 & 7\\
\hline 6 & 1 & 2 & 4 & 7 & 5 & 3 & 8\\
\hline 5 & 2 & 3 & 6 & 4 & 7 & 8 & 1\\
\hline
\end{tabular}
}
}
\caption{Latin squares for singular states with $|h|>1$ and $\measuredangle h \in [0,\pi/4]$ for ${\mathcal{S}}_{8}$} 
\label{tab:LS_8}
\end{figure*}

Due to the symmetry of ${\mathcal{S}}_{8}$ with respect to rotation of $\pi/4$ degrees, the technique used in \cite{NMRarX} can be applied to obtain the Latin squares corresponding to all the 108 singular fade states from those corresponding to the singular fades lying on or outside the unit circle and in the angular interval $\left[ 0, \pi/4 \right]$ which comes out to be 10 in number. These are given in Fig. \ref{tab:LS_8} and Fig. \ref{fig:LS1}.

\begin{figure}[ht]
\centering
\tiny{
\subfigure[$h=1$]{
\begin{tabular}{|c|c|c|c|c|c|c|c|}
\hline 1 & 2 & 3 & 4 & 5 & 6 & 7 & 8\\
\hline 2 & 1 & 4 & 3 & 6 & 5 & 8 & 7\\
\hline 3 & 4 & 1 & 2 & 7 & 8 & 5 & 6\\
\hline 4 & 3 & 2 & 1 & 8 & 7 & 6 & 5\\
\hline 5 & 6 & 7 & 8 & 1 & 2 & 3 & 4\\
\hline 6 & 5 & 8 & 7 & 2 & 1 & 4 & 3\\
\hline 7 & 8 & 5 & 6 & 3 & 4 & 1 & 2\\
\hline 8 & 7 & 6 & 5 & 4 & 3 & 2 & 1\\
\hline
\end{tabular}
}
\subfigure[$h=\frac{\sqrt{3}}{2} + \frac{1}{2} i$]{
\begin{tabular}{|c|c|c|c|c|c|c|c|}
\hline 1 & 8 & 2 & 3 & 7 & 4 & 6 & 5\\
\hline 8 & 7 & 4 & 5 & 2 & 6 & 1 & 3\\
\hline 4 & 6 & 5 & 2 & 3 & 7 & 8 & 1\\
\hline 5 & 4 & 7 & 6 & 1 & 8 & 3 & 2\\
\hline 3 & 5 & 1 & 4 & 8 & 2 & 7 & 6\\
\hline 7 & 3 & 8 & 1 & 6 & 5 & 2 & 4\\
\hline 6 & 2 & 3 & 8 & 5 & 1 & 4 & 7\\
\hline 2 & 1 & 6 & 7 & 4 & 3 & 5 & 8\\
\hline
\end{tabular}
}}
\caption{Latin squares for singular states with $|h|=1$ and $\measuredangle h \in [0,\pi/4]$ for ${\mathcal{S}}_{8}$}
\label{fig:LS1}
\end{figure}

\section{Channel Quantization}
In the physical layer network coding schemes employing Latin Squares to obtain network maps, channel quantization is done to partition the complex plane representing the fade state and to specify which Latin Square is to be used for a generic value of the fade state realization $\gamma e^{j \theta} \in \mathbb{C}$. In this section we derive the channel quantization for ${\mathcal{S}}_{4}$ and ${\mathcal{S}}_{8}$. 

The values of $\gamma e^{j \theta}$ for which any network coding map satisfying the exclusive law gives the same minimum cluster distance value is referred to as the \textit{clustering independent} region. The region in the complex plane other than the clustering independent region is called the \textit{clustering dependent} region. It was shown in \cite{VNR} and \cite{NMRarX} that the clustering independent region, denoted by $\Gamma_{CI}({\cal S})$ consists of those values of $\gamma e^{j \theta}$ which satisfy the following condition,
\begin{align}
\label{eqn:clust_ind}
{d_{min}(\gamma e^{j\theta}) \geq \min \{ d_{min}({\cal S}), \gamma d_{min}({\cal S})\}}.
\end{align}
The region in $\Gamma_{CI}({\cal S})$ where \mbox{$d_{min}(\gamma e^{j\theta}) \geq  d_{min}({\cal S})$} is denoted by $\Gamma_{CI}^{ext}({\cal S})$ and the region where $d_{min}(\gamma e^{j\theta}) \geq  \gamma d_{min}({\cal S})$ is denoted by $\Gamma_{CI}^{int}({\cal S})$. Since $\Gamma_{CI}^{int}({\cal S})$ can be obtained as the complex inversion of $\Gamma_{CI}^{ext}({\cal S})$, finding out one, the other can be easily evaluated \cite{VNR}.

Associated with each singular fade state $z \in {\cal H}$, a region ${\cal R}_{\left\lbrace z\right\rbrace}$ is identified in the clustering dependent region, in which the clustering  ${\cal C}^{\left \{z \right \}}$ is used. The region ${\cal R}_{\left\lbrace z\right\rbrace}$ is obtained essentially by plotting the curves $c(z, z')$, which is referred to as the pair-wise transition boundary between the regions ${\cal R}_{\left\lbrace z\right\rbrace}$ and ${\cal R}_{\left\lbrace z' \right\rbrace}$.

\noindent Let $\check{d_l}$ and $\check{d_{l'}}$ be defined as follows:
\begin{align*}
\check{d_l}={\mathrm{arg}} \hspace{-0.2cm}\min_{d_2 \in \Delta {\cal S}} \left\{ \vert d_1 + z d_2 \vert \right\},~
\check{d_{l'}}={\mathrm{arg}} \hspace{-0.2 cm}\min_{d_2 \in \Delta {\cal S} } \left\{ \vert d_1 + z' d_2 \vert \right\}.
\end{align*}Then the pair wise transition curve $c\left( z, z' \right)$, $z \neq z'$ is any one of the following:
\begin{itemize}
\item if $|\check{d_l}| \neq |\check{d_{l'}}|$, a circle with center $\left( x, y\right)$ and radius $r$, where
\begin{align*}
&x=\frac{\Re e\left( z\right)}{1- {| \frac{\check{d_{l'}}}{\check{d_l}} |}^2} + \frac{\Re e\left( z'\right)}{1- {| \frac{\check{d_l}}{\check{d_{l'}}} |}^2},~
y=\frac{\Im m\left( z\right)}{1- {| \frac{\check{d_{l'}}}{\check{d_l}} |}^2} + \frac{\Im m\left( z'\right)}{1- {| \frac{\check{d_l}}{\check{d_{l'}}} |}^2}\\
&\mathrm{and}~r=\sqrt{\left( x^2 +y^2 + \frac{|z'|^2 |\check{d_{l'}}|^2-|z|^2 |\check{d_{l}}|^2}{|\check{d_{l}}|^2-|\check{d_{l'}}|^2}\right)}
\end{align*}

\item {if $|\check{d_l}| = |\check{d_{l'}}|$, a straight line of the form $ax + by=c$, where \vspace{-0.2 in}\begin{align*} 
a&=\left( {\Re e\left( z\right)|\check{d_l}|^2-\Re e\left( z'\right)|\check{d_{l'}}|^2} \right),\\
b&=\left( {\Im m\left( z\right)|\check{d_l}|^2-\Im m\left( z'\right)|\check{d_{l'}}|^2} \right),\\
c&=-\frac{1}{2} \left( {|z'|^2 |\check{d_{l'}}|^2-|z|^2 |\check{d_{l}}|^2} \right).
\end{align*}}
\end{itemize}

\begin{figure}[htbp]
\centering
\subfigure[Shaded region shows $\Gamma_{CI}^{ext}({{\mathcal{S}}_{4}})$]{
\includegraphics[totalheight=2in,width=2in]{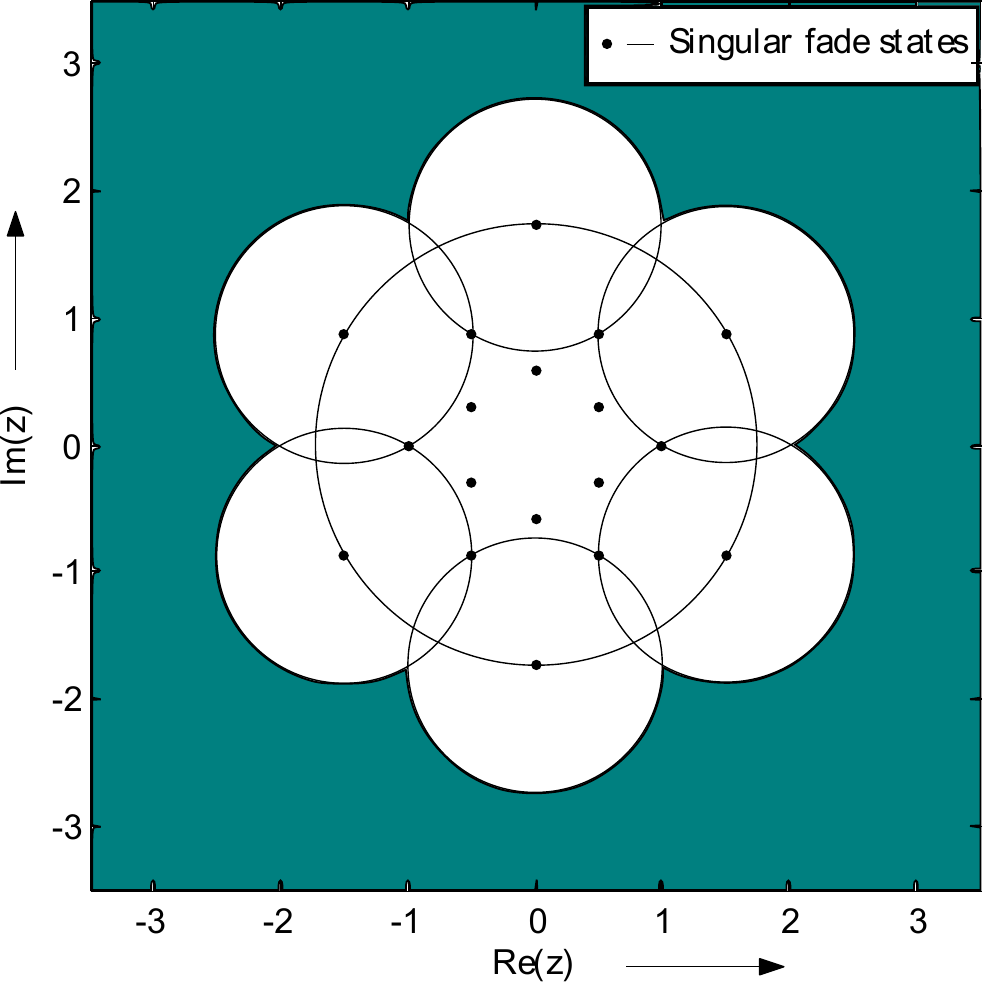}
\label{fig:CIext_4}
}
%\vspace{0.5 cm}
\subfigure[Shaded region shows $\Gamma_{CI}^{int}({{\mathcal{S}}_{4}})$]{
\includegraphics[totalheight=2in,width=2in]{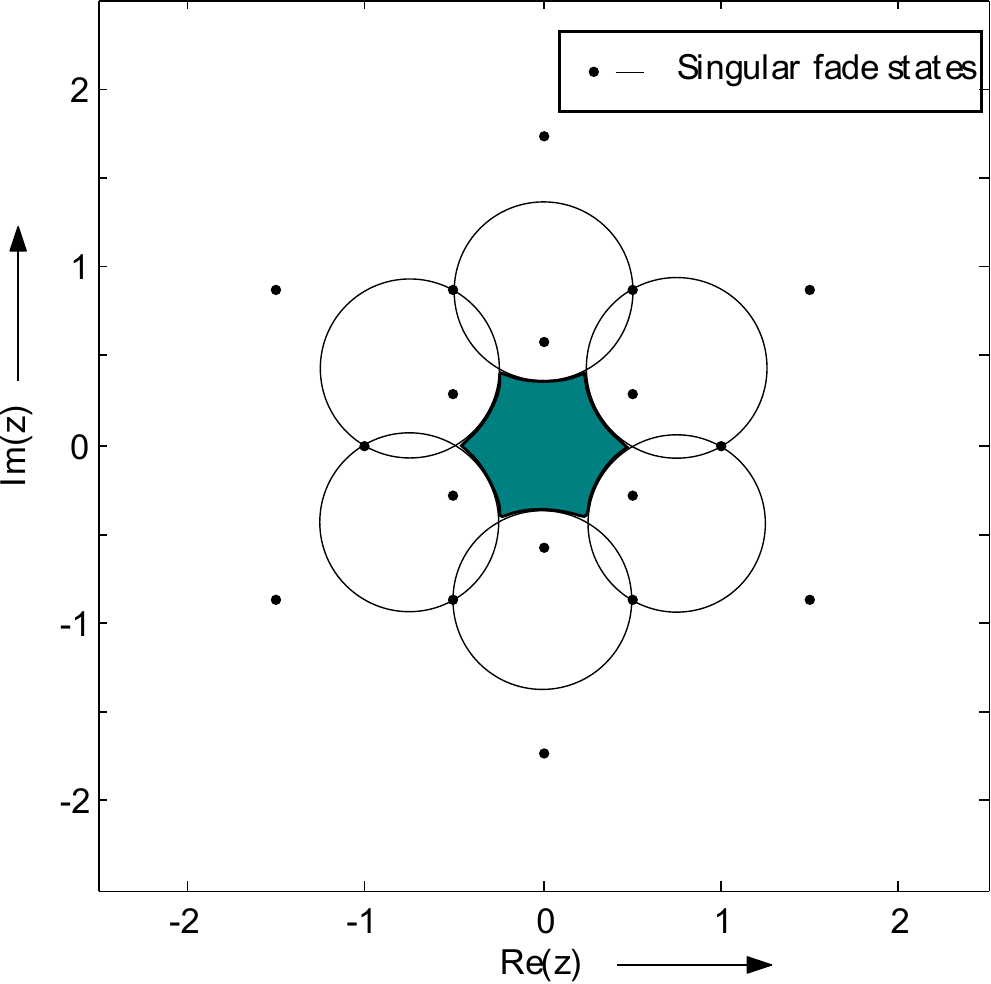}
\label{fig:CIint_4}
}
\caption{Diagram showing $\Gamma_{CI}^{ext}$ and $\Gamma_{CI}^{ext}$ regions for ${\mathcal{S}}_{4}$}
\label{fig:CIextint_4}
\end{figure}

\noindent \textbf{\textit{ Channel Quantization for ${\mathcal{S}}_{4}$}} : Recalling that the singular fade states of ${\mathcal{S}}_{4}$ lie on three concentric circles centered at the origin and with radii $\{ \sqrt{3}, 1, 1/\sqrt{3} \}$, the region $\Gamma_{CI}^{ext}({{\mathcal{S}}_{4}})$ is the exterior of the six unit circles centered at the singular fade states lying on the outermost circle (of radius $\sqrt{3}$). This is obtained by evaluating the expression $d_{min}(\gamma e^{j\theta}) \geq  d_{min}({\cal S})$. The shaded region in Fig. \ref{fig:CIext_4} shows $\Gamma_{CI}^{ext}({{\mathcal{S}}_{4}})$. The region  $\Gamma_{CI}^{int}({{\mathcal{S}}_{4}})$ is obtained by the complex inversion of the region  $\Gamma_{CI}^{ext}({{\mathcal{S}}_{4}})$ and is given by the region lying exterior to the circles of radius $1/2$ and centered at locations $\{ \frac{3}{2}z: |z|=1/\sqrt{3} \}$. $\Gamma_{CI}^{int}({{\mathcal{S}}_{4}})$ is shown as the shaded region in Fig. \ref{fig:CIint_4}. 

\begin{figure}[htbp]
\centering
\includegraphics[totalheight=2in,width=2in]{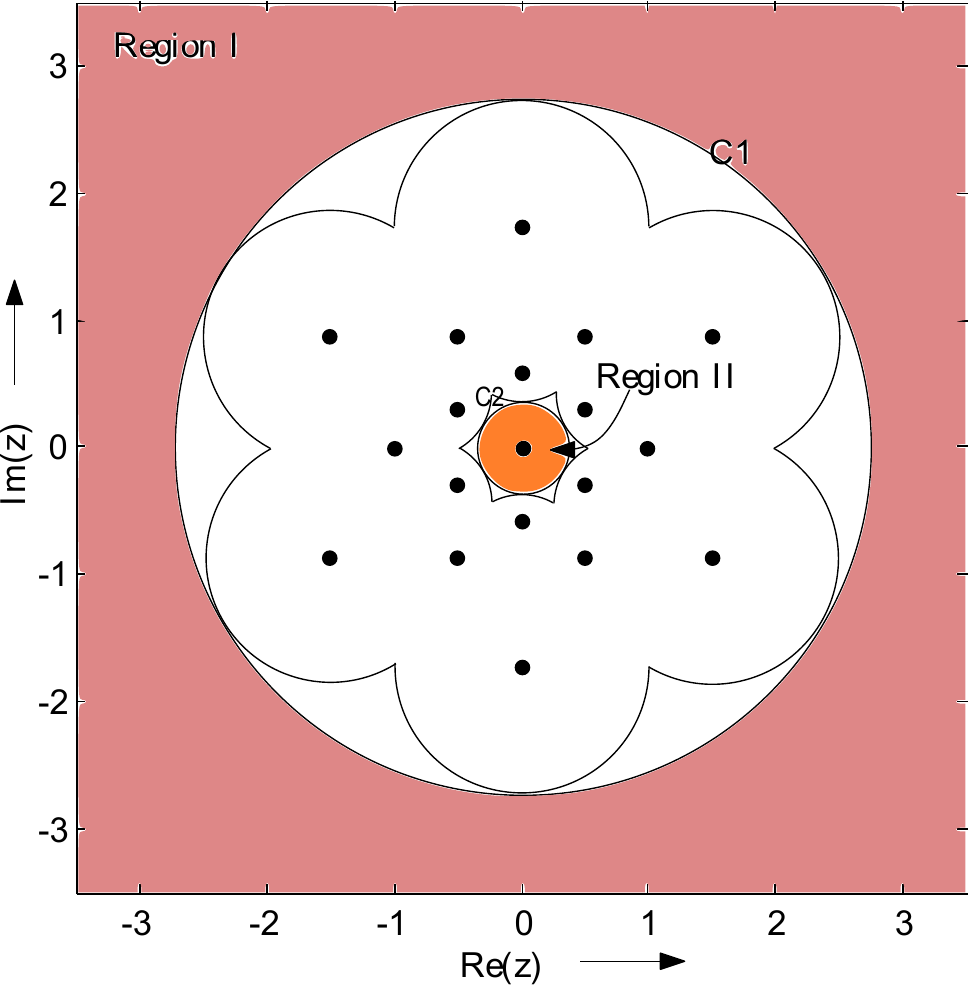} 
\caption{Region I and Region II which can be evaluated based on $|z|$, where relay can use any network coding map for $\mathcal{S}_4$}
\label{fig:adaptive_sw4}
\end{figure}

\begin{figure}[htbp]
\centering
\includegraphics[totalheight=2in,width=2in]{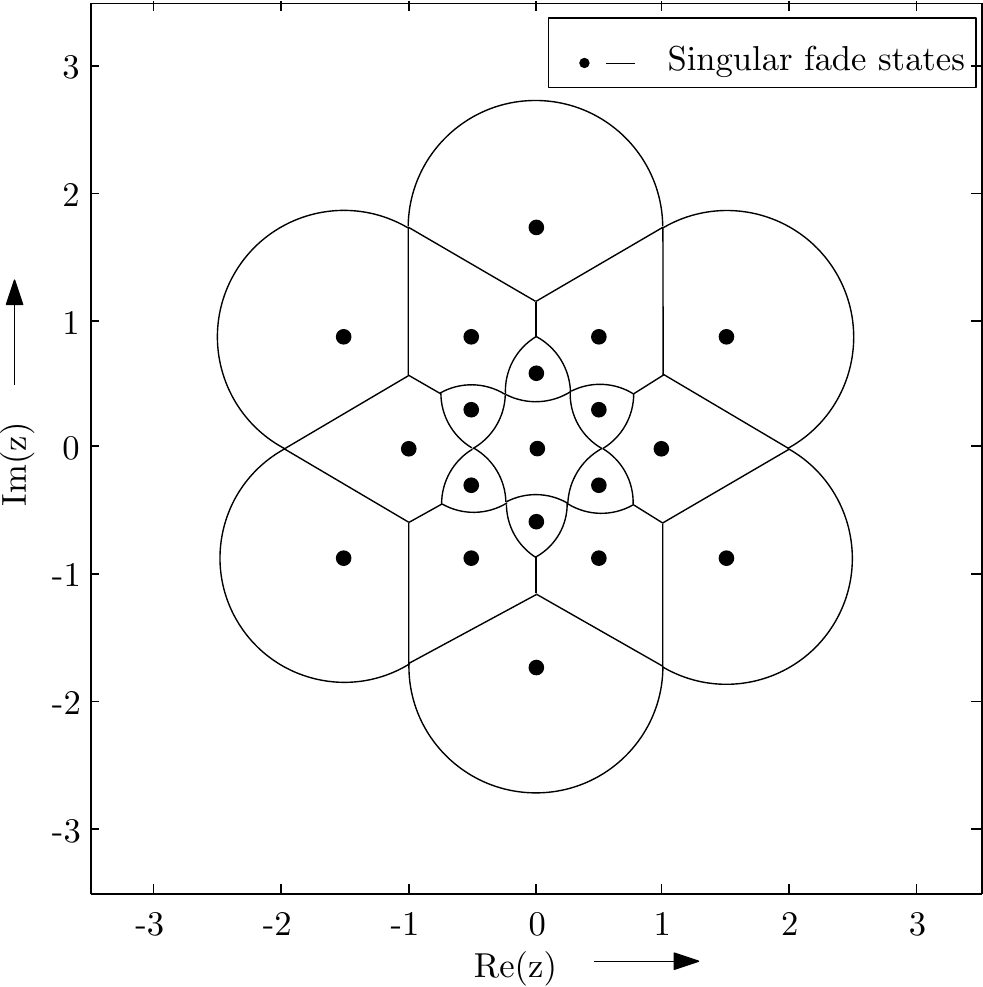} 
\caption{Diagram showing the quantization of the $\gamma e^{j \theta}$ plane for ${\mathcal{S}}_{4}$}
\label{fig:Ch_qn_4}
\end{figure}

As an approximation, in order to reduce the computational complexity at R, the relay can choose to treat all those values of $\gamma e^{j \theta}$, that fall outside the circle of radius $\sqrt{3}+1$ (C1 in Fig. \ref{fig:adaptive_sw4}) or inside the circle of radius $1/(\sqrt{3}+1)$ (C2 in Fig. \ref{fig:adaptive_sw4}), to constitute the Clustering independent region and hence use any network coding map of Fig. \ref{tab:LS_4}. In the clustering dependent region, by plotting the pair-wise transition curves $c\left( z, z' \right)$, $z \neq z'$, we get the complete channel quantization for ${\mathcal{S}}_{4}$ as shown in Fig. \ref{fig:Ch_qn_4}.

\begin{figure}[htbp]
\centering
\subfigure[Shaded region shows $\Gamma_{CI}^{ext}({{\mathcal{S}}_{8}})$]{
\includegraphics[totalheight=2in,width=2in]{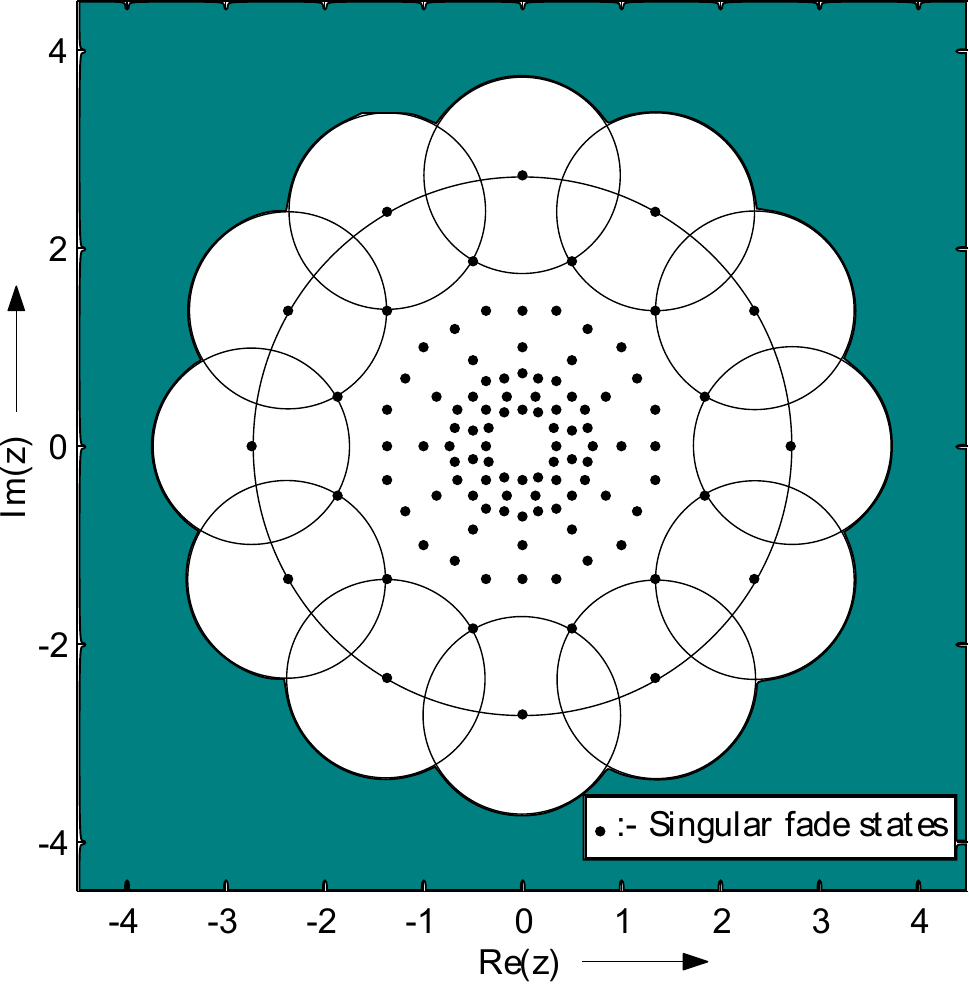}
\label{fig:CIext_8}
}
%\vspace{0.5 cm}
\subfigure[Shaded region shows $\Gamma_{CI}^{int}({{\mathcal{S}}_{8}})$]{
\includegraphics[totalheight=2in,width=2in]{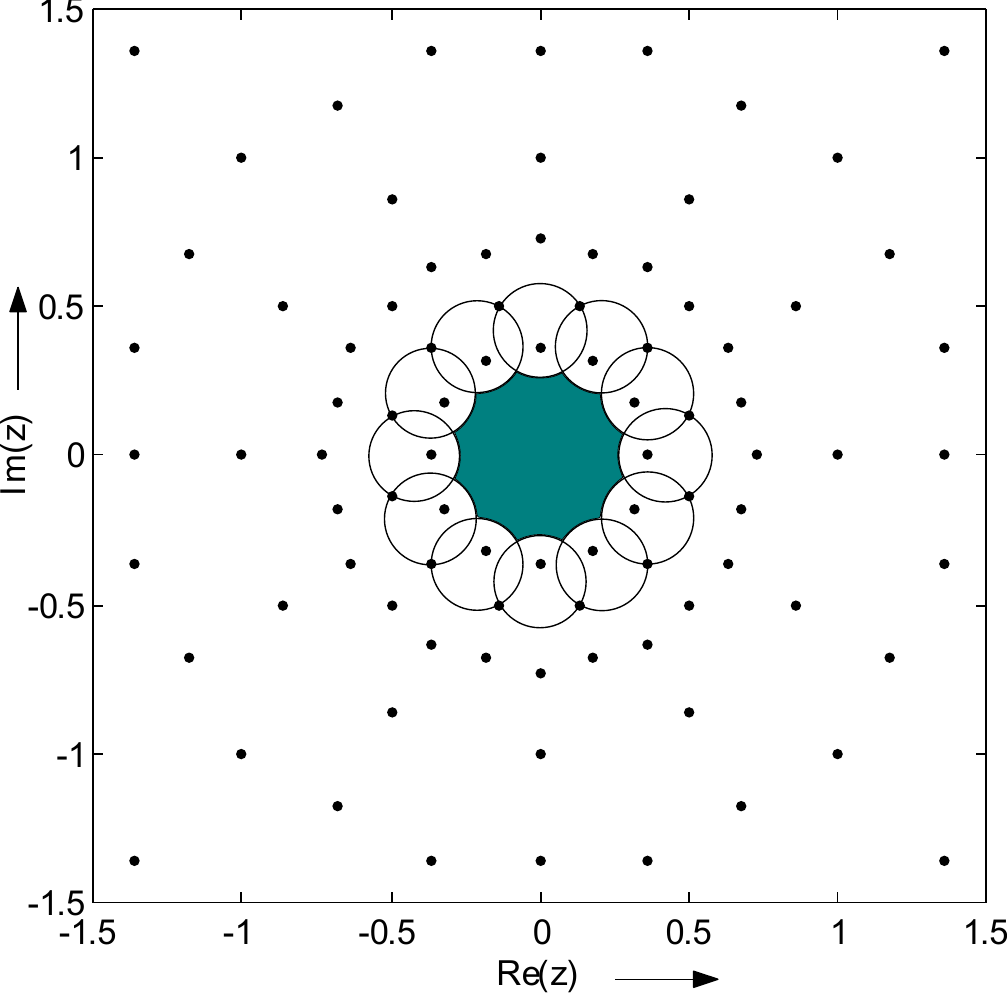}
\label{fig:CIint_8}
}
\caption{Diagram showing $\Gamma_{CI}^{ext}$ and $\Gamma_{CI}^{ext}$ regions for ${\mathcal{S}}_{8}$}
\label{fig:CIextint_8}
\end{figure}

\begin{figure}[htbp]
\centering
\includegraphics[totalheight=2in,width=2in]{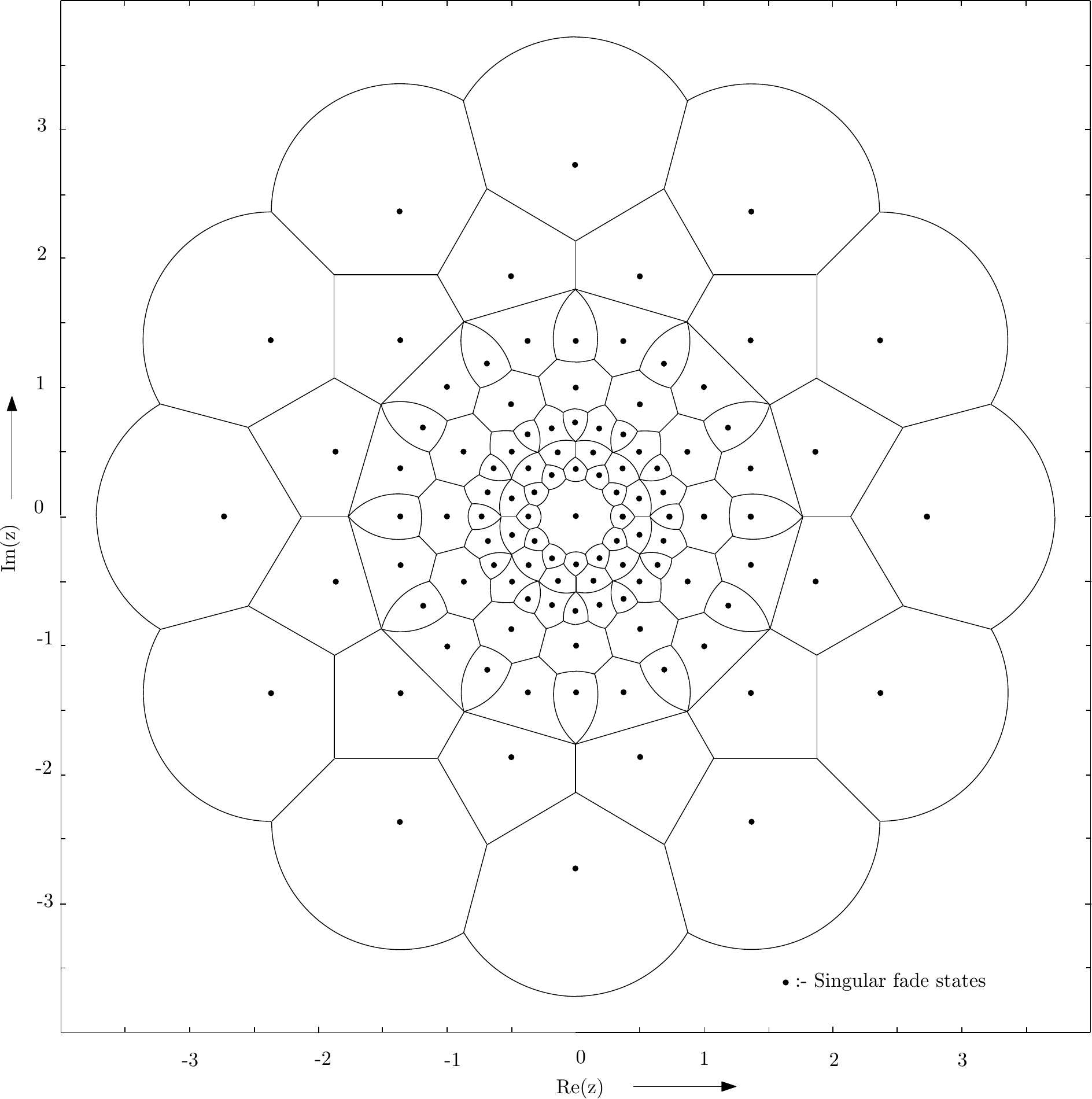} 
\caption{Diagram showing the quantization of the $\gamma e^{j \theta}$ plane for ${\mathcal{S}}_{8}$}
\label{fig:Ch_qn_8}
\end{figure}

\vspace{0.1 in}
\noindent \textbf{\textit{ Channel Quantization for ${\mathcal{S}}_{8}$}} : First notice that the maximum and the minimum values for $|z|, z \in {\mathcal{H}}$ for ${\mathcal{S}}_{8}$ are $1+\sqrt{3}$ and $1/(1+\sqrt{3})$ respectively. The region $\Gamma_{CI}^{ext}({{\mathcal{S}}_{8}})$ is the exterior of the unit circles centered at the singular fade states $z$ with $|z| = 1+\sqrt{3}$. Fig. \ref{fig:CIext_8} shows the region $\Gamma_{CI}^{ext}({{\mathcal{S}}_{8}})$ and the region $\Gamma_{CI}^{int}({{\mathcal{S}}_{8}})$ obtained as the complex inversion of $\Gamma_{CI}^{ext}({{\mathcal{S}}_{8}})$ is shown in Fig. \ref{fig:CIint_8}. Finally, the complete channel quantization for ${\mathcal{S}}_{8}$ is shown in Fig. \ref{fig:Ch_qn_8}.

\begin{figure}[htbp]
\centering
\includegraphics[totalheight=2in,width=2in]{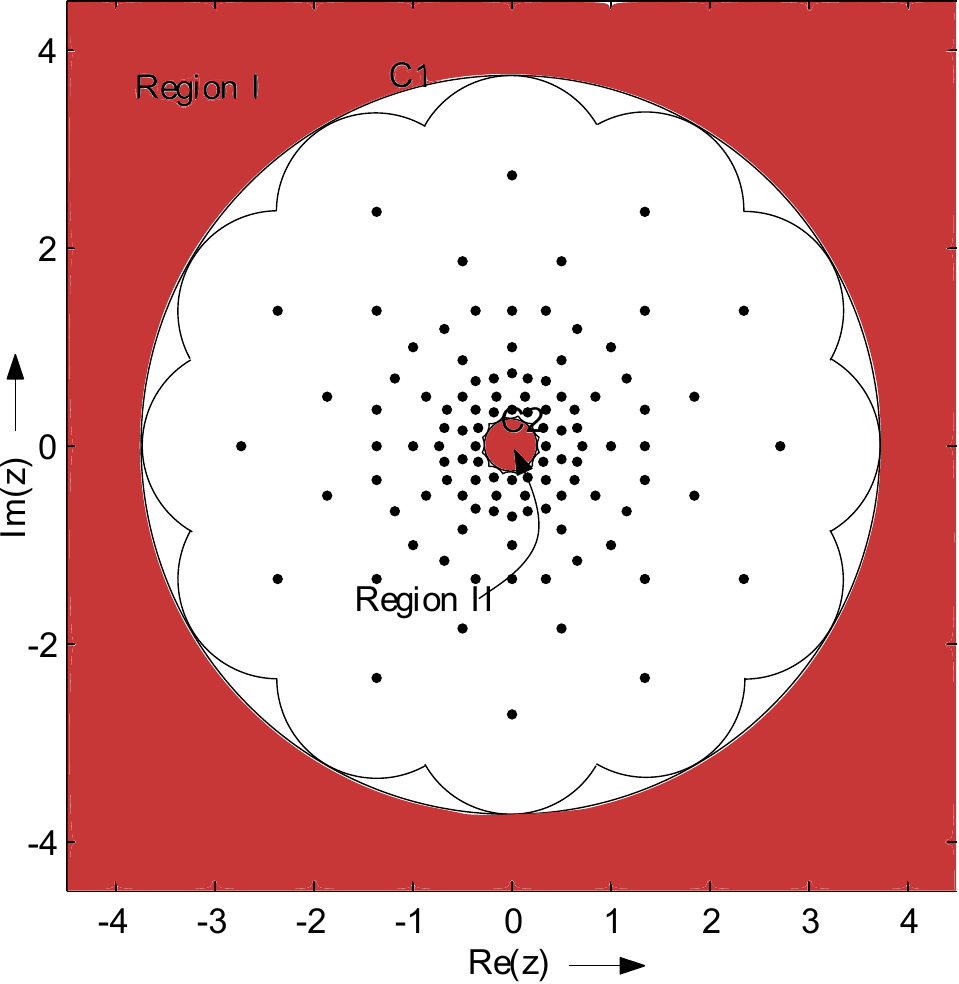} 
\caption{Region I and Region II which can be evaluate based on $|z|$, where relay can use any network coding map for $\mathcal{S}_8$}
\label{fig:adaptive_sw8}
\end{figure}

The relay can choose to treat all those values of $\gamma e^{j \theta}$, that fall outside the circle of radius $\sqrt{3}+2$ (C1 in Fig. \ref{fig:adaptive_sw8}) or inside the circle of radius $2-\sqrt{3})$ (C2 in Fig. \ref{fig:adaptive_sw8}), to constitute the clustering independent region and hence use any network coding map of Fig. \ref{tab:LS_8}. The relay needs to adaptively switch between the network coding maps only if the fade state value falls in the region between the circles C1 and C2 in Fig. \ref{fig:adaptive_sw8}. This reduces the computational complexity incurred at R.

\section{Simulation Results}
\label{sec5}
\begin{figure*}
\centering
\includegraphics[totalheight=4.2in,width=6in]{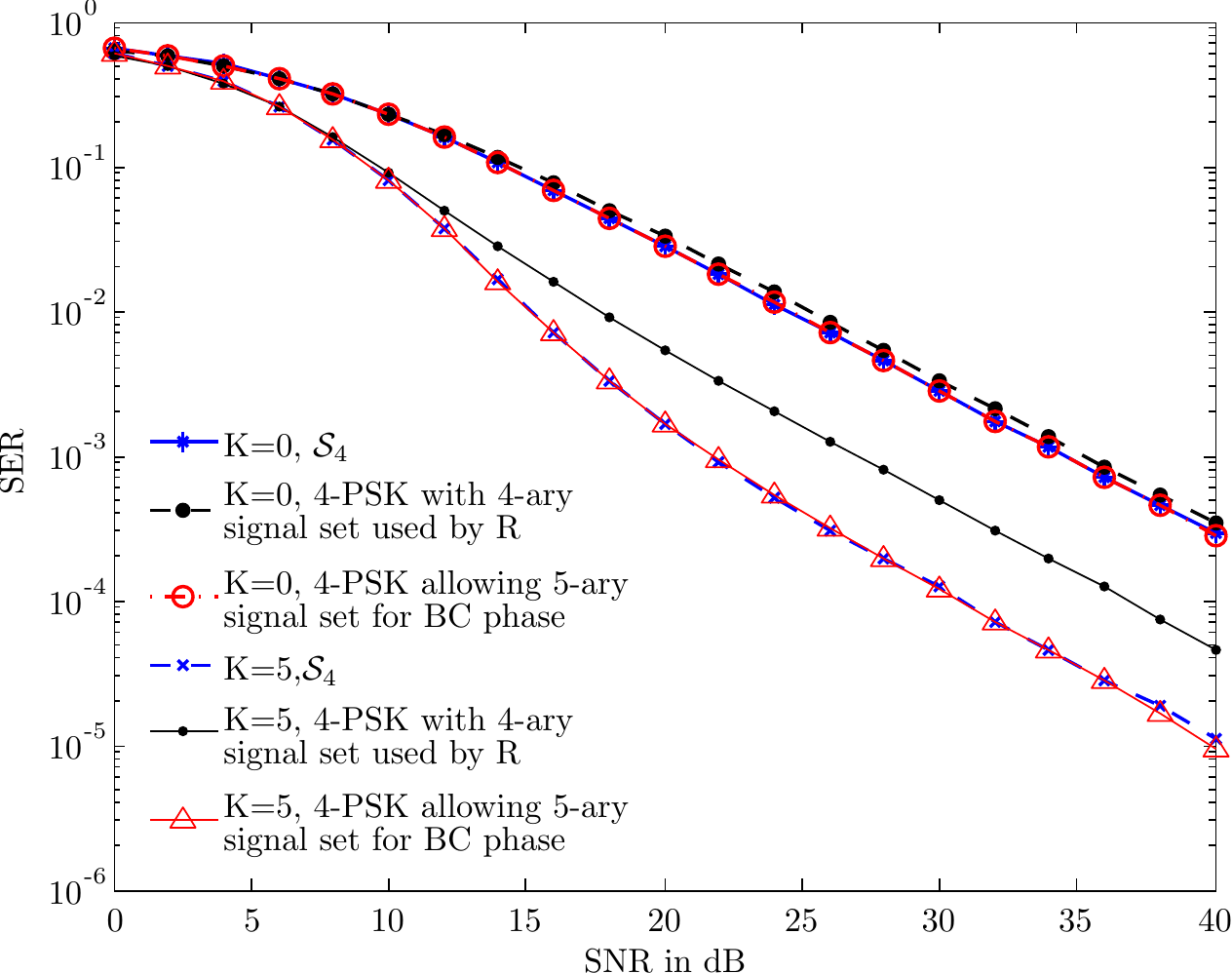} 
\caption{SNR vs Average SER plots for ${\mathcal{S}}_{4}$ and 4-PSK for Rician factor K=0 and K=5 when R is allowed to use only 4-ary constellation and otherwise}
\label{fig:ser4}
\end{figure*}

In this section we present the simulation results showing the average end-to-end SER for the proposed constellations ${\mathcal{S}}_{4}$ and ${\mathcal{S}}_{8}$ and compare them with those for 4-PSK and 8-PSK under different fading scenarios. First consider the case when the end nodes A and B use 4-point constellations. The average SER curves as a function of SNR in dB for ${\mathcal{S}}_{4}$ and 4-PSK when the fade coefficients $h_A, h_B, h'_A$ and $h'_B$ are Rayleigh distributed (Rician fading with K=0) and when the fade coefficients are Rician distributed with Rician factor K=5 are given in Fig. \ref{fig:ser4}. Fig \ref{fig:ser4} also shows the performance of 4-PSK signal set when the relay R uses only 4-point constellation for BC phase. In this scenario ${\mathcal{S}}_{4}$ gives roughly 1 dB advantage over 4-PSK. The same performance as that given by ${\mathcal{S}}_{4}$ can be achieved by 4-PSK only if the relay is allowed to use 5-point constellation. It may be noted here that there is no degradation of performance for ${\mathcal{S}}_{4}$ due to the fact that it has more number of singular fade states than 4-PSK.

\begin{figure*}
\centering
\includegraphics[totalheight=4.2in,width=6in]{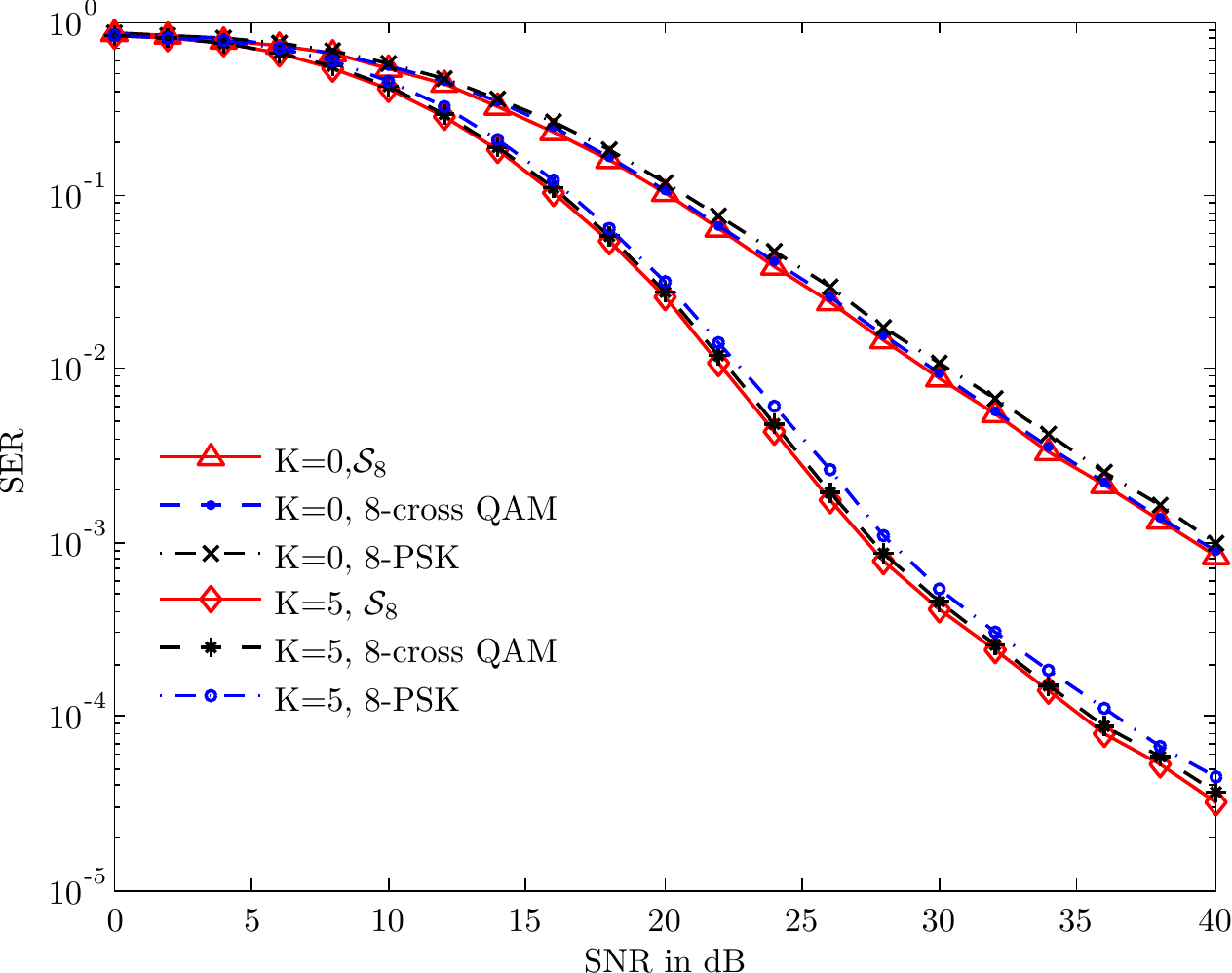} 
\caption{SNR vs Average SER plots for ${\mathcal{S}}_{8}$, 8-cross QAM and 8-PSK for Rician factor K=0 and K=5}
\label{fig:ser8}
\end{figure*}

When the end nodes A and B use 8-point constellations, the average SER vs SNR curves for ${\mathcal{S}}_{8}$, 8-cross QAM and 8-PSK are given in Fig. \ref{fig:ser8} for the case when the fade coefficients $h_A, h_B, h'_A$ and $h'_B$ are Rayleigh distributed (Rician fading with K=0) and also when K=5. Table \ref{8_point_ compare} shows that 8-PSK has the least number of singular fade states among the three. However, from Fig. \ref{fig:ser8}, it is seen that 8-PSK has the worst SER performance. So, the number of singular fade states has no effect on the SER performance. It is seen that ${\mathcal{S}}_{8}$ gives 0.8 dB better average end-to-end SER performance than 8-PSK for Rayleigh fading scenario and also outperforms 8-cross QAM. For Rician fading scenario with Rician factor K=5, ${\mathcal{S}}_{8}$ gives 1.7 dB better average end-to-end SER performance than 8-PSK. Hence ${\mathcal{S}}_{8}$ is the best choice among $M$=8 signal sets.

\section{Discussion}
In this paper we discuss the desirable features in the signal set used during the MA phase for bi-directional wireless relaying using physical layer network coding. 4-PSK requires the use of 5-ary constellation at R for several fade states. We propose a 4-point constellation that always requires a 4-point constellation during the BC phase and whose SER performance can be achieved by 4-PSK only if the relay is allowed to use a 5-ary constellation. We show that the number of singular fade states does not play any role in determining the end-to-end SER performance in bi-directional wireless relaying. We propose an 8-point signal set that always performs better than 8-PSK. Not only does it require 8-point constellation for the BC phase like 8-PSK for all channel realisations, but also has just four singular fade states more than that of 8-PSK. Further  studies can focus on the construction of signal sets giving best end-to-end SER performance for arbitrary values of $M$.
%%%%%%%%%%%%%%%%%%%%%%%%%%%%%%%%%%%%%%%%%%%%%%%%%%%%%%%%%%%%%%%%%%%%%%%%%%%%%%%%%%%%%%

%\bibliographystyle{IEEEtran.bst}
\bibliography{IEEEabrv,3rdsem_bib}
\end{document}